\documentclass[10pt,twocolumn,letterpaper]{article}
\usepackage[pagenumbers]{cvpr} %

\usepackage{multirow}
\usepackage{titletoc}
\usepackage{xspace}
\newcommand{\methodname}{\textsc{DiffSimp}\xspace}

\usepackage{amsthm}

\usepackage{placeins}

\definecolor{cvprblue}{rgb}{0.21,0.49,0.74}
\usepackage[pagebackref,breaklinks,colorlinks,allcolors=cvprblue]{hyperref}

\def\confName{CVPR}
\def\confYear{2027}

\title{DiffSimp: Differentiable Mesh Simplification for\texorpdfstring{\\}{ } Gradient-Based Low-Poly Modeling}

\author{Seonghun Oh\\
Holiday Robotics\\
{\tt\small seonghun.oh@holiday-robotics.com}
}

\begin{document}
\twocolumn[{%
\renewcommand\twocolumn[1][]{#1}%
\maketitle
}]
\begin{abstract}
Differentiable mesh extraction enables
gradient-based optimization of detailed 3D geometry
across diverse tasks.
However, compact meshes are commonly obtained
through post-processing simplification, leaving
simplification-induced errors outside upstream
optimization.
We introduce \methodname, a training-free, drop-in
differentiable mesh simplification layer built on
QEM-based parallel edge collapse, enabling existing
geometry optimization pipelines to optimize
their final simplified meshes.
We formulate simplification under a fixed
collapse history as a locally differentiable
collapse map, enabling gradient propagation
without relaxing discrete collapse decisions.
We combine fast, high-quality parallel
simplification with a factorized backward
formulation that exploits differentiable vertex
quadrics and quadric reduction forests for
efficient gradient propagation.
Experiments on low-poly mesh reconstruction and
inverse rendering demonstrate improved fidelity
over post-processing simplification, with additional
gains from gradient propagation that increase
under tighter vertex budgets.
\methodname supports diverse geometry optimization
frontends, with 1.00--1.25$\times$ the end-to-end
runtime of unsimplified baselines in evaluated
pipelines.
\end{abstract}

\section{Introduction}
\label{sec:intro}

Triangle meshes remain a fundamental surface representation
across computer graphics, 3D vision, and robotics.
Recent advances in differentiable mesh extraction~\cite{remelli2020meshsdf,shen2021deep,
shen2023flexible,binninger2025tetweave,
wei2025neumanifold}
have enabled gradient-based geometry optimization
by propagating gradients from mesh-based objectives
upstream, supporting
inverse rendering~\cite{munkberg2022extracting,hasselgren2022nvdiffrecmc,
ye2025geosplatting},
scene reconstruction~\cite{guedon2025milo,cui2026sdfraster},
and generative 3D modeling~\cite{gao2022get3d,xiang2024structured,
sam3dteam2025sam3d3dfyimages}.
However, these pipelines often produce dense meshes
to capture fine geometric structures, while downstream
applications may require compact meshes.

A straightforward way to obtain compact meshes
is to reduce the resolution of differentiable
mesh extraction, but this limits the recovery
of fine geometric details during optimization.
Instead, a common strategy is to reconstruct
or generate sufficiently detailed meshes,
followed by post-processing simplification
to meet downstream complexity requirements.
Among mesh simplification methods,
quadric error metric (QEM) edge collapse~\cite{garland1997surface,garland1999quadric}
has become a practical off-the-shelf choice~\cite{knodt2025single,liu2025simplifying}.
It naturally adapts mesh complexity to local
geometry, enabling aggressive simplification
while limiting geometric approximation error,
without requiring separate iterative fitting
as in variational approaches~\cite{cohensteiner2004variational,xu2024cwf}
or pretrained simplification models~\cite{potamias2022neural}.
These advantages have made QEM a common
post-processing step in modern 3D pipelines~\cite{hunyuan3d2025hunyuan3d,
wang2026assetgen,choi2024ltm, gao2025efficient}.

However, post-processing mesh simplification
decouples geometry optimization from the final
simplified mesh.
While differentiable mesh extraction enables
gradient propagation from mesh-based objectives
upstream, these objectives are typically evaluated
on detailed intermediate meshes rather than
the simplified meshes ultimately deployed.
This mismatch becomes particularly consequential
under aggressive simplification, where even QEM
can introduce substantial geometric distortion
as fine structures are approximated by
increasingly coarse meshes.
These simplification-induced errors provide
no gradient signal to the upstream optimizer,
preventing it from directly compensating
for them to improve the final deployed mesh.

To address this optimization--deployment mismatch,
we seek to integrate mesh simplification as a
differentiable layer into gradient-based
geometry optimization pipelines.
For practical integration, we identify three
design priorities.
First, the layer should be training-free and
mesh-native, independent of learned simplification
priors and specific upstream representations,
enabling its use across diverse pipelines
and downstream objectives.
Second, high forward simplification quality must
be maintained.
At less restrictive budgets, simplification
typically introduces smaller geometric errors,
leaving limited room for gradient-based compensation
and making forward quality particularly important.
Under tighter budgets, high-quality forward
simplification and gradient-based compensation
become complementary in preserving geometric fidelity.
Third, efficient forward and backward execution
is essential, as simplification must be performed
at every optimization iteration.

To meet these priorities, we introduce
\emph{\methodname}, a training-free, drop-in
differentiable mesh simplification layer
built on QEM-based parallel edge collapse, enabling existing
gradient-based geometry optimization pipelines
to optimize through their simplified meshes.
Although discrete edge-collapse decisions prevent
global differentiability, we exploit the recorded
collapse history to express simplification as a
locally differentiable collapse map, enabling
gradient propagation without relaxing these decisions.
To achieve fast, high-quality forward simplification,
we extend the parallel edge-collapse scheme of
PaMO~\cite{oh2025pamo}, improving simplification
quality while retaining efficient execution.
For efficient backward propagation, we propose
a factorized differentiation scheme that treats
vertex quadrics as differentiable intermediates,
enabling parallel collapse-local vector--Jacobian
products (VJPs), and exploits the quadric reduction-forest
structure to avoid redundant face-level differentiation.
We implement both operations in CUDA and introduce
a progressive vertex-budget schedule for integration
into iterative optimization.

Experiments on low-poly mesh reconstruction and
inverse rendering demonstrate that optimizing
through \methodname improves the fidelity of
final simplified meshes over post-processing
in both tasks.
We also show that gradient propagation provides
additional gains that increase under tighter
vertex budgets, as upstream optimization compensates
for simplification-induced errors.
We further demonstrate compatibility with diverse
existing gradient-based geometry optimization
pipelines.
These improvements are achieved with only
1.00--1.25$\times$ the end-to-end optimization
time of the corresponding unsimplified pipelines.

\section{Related Work}
\label{sec:related_work}

\subsection{Gradient-Based Mesh Optimization}

Differentiable mesh extraction~\cite{remelli2020meshsdf,guillard2024deepmesh,
guillard2022meshudf,shen2021deep,
shen2023flexible,wei2025neumanifold,
binninger2025tetweave}
and differentiable mesh representations~\cite{rakotosaona2021differentiable,
son2024dmesh,son2025dmeshpp}
enable upstream gradient propagation from
mesh-based objectives, supporting diverse
downstream applications~\cite{munkberg2022extracting,ye2025geosplatting,
nicolet2021large,guedon2025milo,cui2026sdfraster,
gao2022get3d,xiang2024structured}.
However, extracted mesh complexity remains tied to
the underlying extraction structure.
Even adaptive extraction methods~\cite{shen2023flexible,binninger2025tetweave}
adjust spatial sampling and extraction structures
rather than performing aggressive,
geometry-adaptive mesh simplification.
Gradient-based low-poly mesh optimization has also
been explored through geometry and appearance
refinement~\cite{hasselgren2021appearance,choi2024ltm},
joint base-mesh and displacement optimization~\cite{dou2024differentiable},
and adaptive remeshing or topology updates~\cite{palfinger2022continuous,gao2026inverse,
fan2026exmesh}.
However, these approaches either optimize compact
meshes from given or reconstructed geometry without
propagating gradients through simplification to
upstream representations, or adapt mesh connectivity
within dedicated optimization pipelines.
In contrast, our reusable differentiable mesh
simplification layer propagates gradients from
objectives on aggressively simplified meshes
to upstream representations without redesigning
existing optimization pipelines.

\subsection{Mesh Simplification}

Mesh simplification has been extensively studied
through vertex clustering~\cite{rossignac1993multiresolution},
variational approximation~\cite{cohensteiner2004variational,
xu2024cwf},
remeshing~\cite{valette2004approximated,
valette2008generic,chen2023robust},
and iterative edge collapse~\cite{hoppe1996progressive,lindstrom1998fast}.
These training-free, mesh-native methods are
widely used as post-processing in modern
3D pipelines.
Among these, quadric error metric (QEM)
edge collapse~\cite{garland1997surface,garland1999quadric}
has become a standard approach to mesh simplification.
To overcome the computational cost of sequential
greedy edge collapse, parallel methods~\cite{papageorgiou2015triangular,oh2025pamo}
accelerate simplification while maintaining
competitive geometric quality, albeit with
some loss relative to sequential QEM.
However, these methods do not propagate gradients
through collapse operations, leaving simplification
largely as post-processing in gradient-based pipelines.
Building on PaMO~\cite{oh2025pamo}, we combine
the geometric adaptivity of QEM and the efficiency
of parallel execution with fixed-history differentiation,
enabling their use in gradient-based optimization.
We further improve parallel simplification quality
to better preserve geometric fidelity under
aggressive simplification while retaining
efficient execution.

\subsection{Learning-Based Low-Poly Modeling}

Learning-based approaches to mesh reduction
and low-poly modeling include task-driven mesh pooling~\cite{hanocka2019meshcnn},
learned simplification operators~\cite{potamias2022neural},
learned geometric representations~\cite{maruani2024ponq},
progressive mesh representations~\cite{chen2024nepm,chen2023neuralprogressive,
lei2025armesh},
and generative mesh priors~\cite{li2026triflow, siddiqui2024meshgpt,chen2025meshanything,
long2026lato2,song2025tssr}.
While these approaches support mesh reduction
and low-poly modeling, they also depend
on learned parameters and training distributions.
Notably, Neural Mesh Simplification~\cite{potamias2022neural}
enables differentiable simplification through
learned vertex sampling and connectivity prediction,
but remains dependent on a pretrained network.
In contrast, our training-free differentiable
layer retains the geometric adaptivity of QEM
without relying on learned simplification priors.

\begin{figure*}[t]
    \centering
    \includegraphics[width=\textwidth]{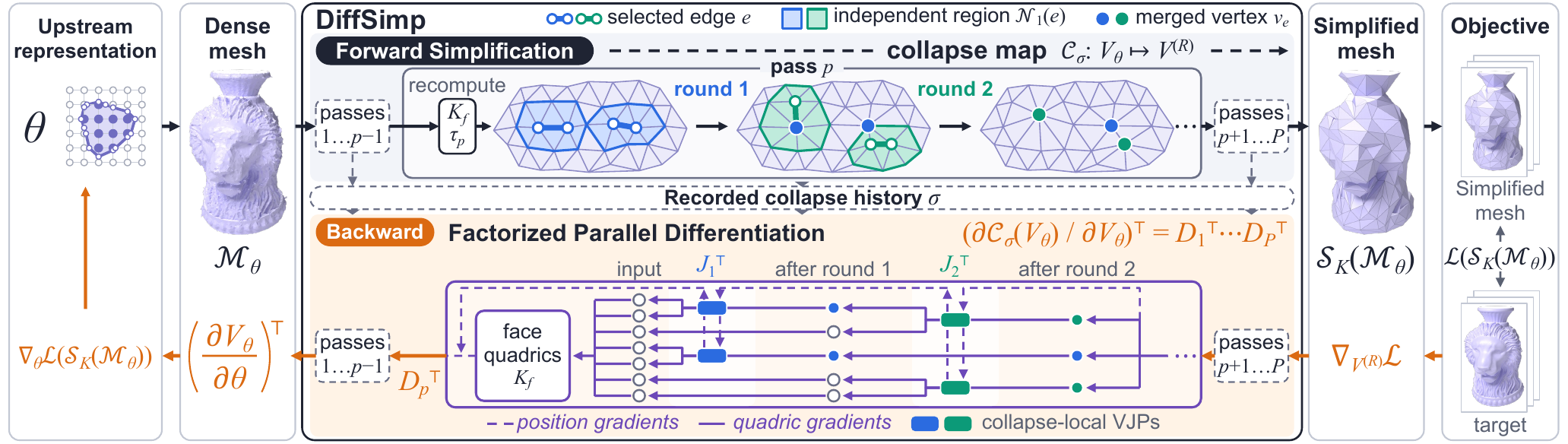}
\caption{
\textbf{Overview of \methodname.}
Given a dense mesh
$\mathcal M_\theta=(V_\theta,F_\theta)$,
\methodname propagates gradients from objectives
on the simplified mesh $\mathcal S_K(\mathcal M_\theta)$
to the upstream parameters $\theta$ in a
gradient-based geometry optimization pipeline.
\textbf{Forward (top):}
Parallel edge-collapse rounds, progressive
cost thresholds ($\tau_p$), and pass-wise quadric
recomputation produce the simplified mesh
while recording the collapse history $\sigma$
(Sec.~\ref{sec:method:forward}).
\textbf{Backward (bottom):}
With $\sigma$ fixed, we differentiate
the collapse map $\mathcal C_\sigma$
(Sec.~\ref{sec:method:fixed_history}).
For each pass, $D_p^\top$ is computed
through factorized parallel differentiation
(Sec.~\ref{sec:method:parallel}).
This combines parallel collapse-local VJPs
within each round (forming $J_r^\top$)
with quadric gradient propagation through
the reduction forest, allowing face quadrics
$K_f$ to be differentiated only once per pass.
}
    \label{fig:overview}
\end{figure*}

\section{Preliminaries}

\subsection{Quadric Error Metrics for Edge Collapse}
\label{sec:prelim_qem}

Quadric error metric (QEM) simplification~\cite{garland1997surface,garland1999quadric}
iteratively simplifies a triangle mesh
through successive edge collapses.
Let $\mathcal{M}^{(0)}=(V^{(0)},F^{(0)})$
denote an input triangle mesh with vertices
$V^{(0)}=\{v_i^{(0)}\}_{i=1}^{N}$,
$v_i^{(0)}\in\mathbb{R}^3$,
and faces $F^{(0)}$.
The simplification proceeds as
$\mathcal{M}^{(0)}
\xrightarrow{\mathcal{E}_1}\cdots
\xrightarrow{\mathcal{E}_R}
\mathcal{M}^{(R)}$,
where $\mathcal{E}_r$ denotes the set of edges
collapsed in round $r$.
In classical sequential QEM, each round greedily collapses the minimum-cost
edge $e=(i,j)$, merging its endpoints $v_i$
and $v_j$ into $v_e$, updating mesh
connectivity and neighboring edge costs.

Before the first round, each face
$f\in F^{(0)}$ defines a face quadric
$K_f=w_fp_fp_f^\top\in\mathbb{R}^{4\times4}$,
where $p_f=(n_f^\top,d_f)^\top$ represents
the supporting plane of face $f$,
with unit normal $n_f$ and offset $d_f$,
and $w_f$ is its weight.
Each vertex $v_i$ is assigned
a vertex quadric
$Q_i=\sum_{f\ni i}K_f$
over its incident faces.
To evaluate the collapse error for an edge
$e=(i,j)$, we first combine the vertex
quadrics of its endpoints:
\begin{equation}
\label{eq:qem_merge}
 Q_e=Q_i+Q_j
=\begin{bmatrix}
A_e & b_e\\
b_e^\top & c_e
\end{bmatrix},
\end{equation}
where $A_e\in\mathbb{R}^{3\times3}$,
$b_e\in\mathbb{R}^3$, and $c_e\in\mathbb{R}$.
The collapse error at position $v$ is
$E_e(v)=\tilde v^\top Q_e\tilde v$,
where $\tilde v=(v^\top,1)^\top$.
When $A_e$ is sufficiently well-conditioned,
the optimal placement is
$v_e=\arg\min_{v\in\mathbb{R}^3}E_e(v)
=-A_e^{-1}b_e$;
otherwise, an edge-segment minimizer or
midpoint fallback is used.
The collapse cost is $C_e=E_e(v_e)$.
Collapsing edge $e$ then updates its
endpoint positions and vertex quadrics as
$(v_i,Q_i),(v_j,Q_j)\mapsto
(v_e,Q_e)$.

\subsection{Parallel Edge Collapse}
\label{sec:prelim_parallel}

Parallel edge collapse contracts multiple
non-conflicting edges in a single round.
PaMO~\cite{oh2025pamo} enables this through
independent-region selection.
For each edge $e=(i,j)$, let
$\mathcal{N}_1(e)=
\mathcal{N}_1(v_i)\cup\mathcal{N}_1(v_j)$
denote its 1-ring face neighborhood, where
$\mathcal{N}_1(v_i)$ contains the faces
incident to $v_i$.
Each edge propagates its collapse cost $C_e$
to all faces in $\mathcal{N}_1(e)$,
and each face selects a single minimum-cost
candidate.
An edge $e$ is selected if it is the
minimum-cost candidate at every face
in $\mathcal{N}_1(e)$.
Since collapsing $e$ only modifies mesh
geometry and topology within $\mathcal{N}_1(e)$,
this neighborhood defines its independent region.
As each face selects a single candidate,
the independent regions
are face-disjoint, enabling conflict-free
parallel collapses.
In each round, selected edges satisfying
$C_e\leq\tau$ are collapsed simultaneously,
where $\tau$ is the cost threshold.
While improving efficiency, parallel execution
may compromise approximation quality relative
to sequential QEM simplification.

\section{Method}
\label{sec:method}

We introduce \methodname (Fig.~\ref{fig:overview}), a differentiable mesh
simplification layer built on QEM-based parallel edge collapse.
Given an upstream mesh generator parameterized by
$\theta$, let $\mathcal{M}_\theta =
(V_\theta,F_\theta)$ denote its output mesh.
We formulate the optimization objective as
\begin{equation}
    \min_{\theta}\;
    \mathcal{L}\left(
        \mathcal{S}_K(\mathcal{M}_\theta)
    \right),
    \label{eq:optimization_objective}
\end{equation}
where $\mathcal{S}_K$ denotes the forward
simplification operator of \methodname
with a target vertex budget $K$.

\subsection{Forward Simplification}
\label{sec:method:forward}

For forward simplification, we extend
the parallel edge-collapse scheme in
Sec.~\ref{sec:prelim_parallel},
using the QEM placement, collapse cost,
and quadric update rules from
Sec.~\ref{sec:prelim_qem}.
To improve simplification quality while
retaining the efficiency of parallel execution,
we progressively increase the cost threshold
across simplification passes until reaching
the target vertex budget $K$.
Each pass $p$ performs parallel collapse rounds
under a cost threshold $\tau_p$ until no further
edges can be collapsed. If the target budget
$K$ is not reached, the threshold is increased
to $\tau_{p+1}$.
Following memoryless simplification
\cite{lindstrom1998fast,hoppe1999new},
we recompute face and vertex quadrics
at the beginning of each pass
to reflect updated mesh geometry.
We also use area-weighted face quadrics
$K_f=w_fp_fp_f^\top$, with
$w_f=a_f/\bar a$, where $a_f$ is the
face area and $\bar a$ is the mean
face area at the beginning of each pass.\footnote{The normalization by $\bar a$ is omitted
from differentiation, as uniform quadric
scaling does not affect vertex placement.}
For numerical stability, we use the regularized
QEM placement
$v_e=-(A_e+\lambda_e I)^{-1}b_e$
for well-conditioned $A_e$, where
$\lambda_e=\eta\max_k |(A_e)_{kk}|$
and $\eta$ is a small coefficient.
Otherwise, an edge-segment or
midpoint fallback is used, yielding
three possible placement branches.

\subsection{Fixed-History Differentiation}
\label{sec:method:fixed_history}

Edge collapse involves discrete edge
selections and connectivity updates,
making the overall simplification process
not globally differentiable.
However, given a fixed collapse history,
the output connectivity is determined,
while the output vertex positions are
obtained from the input vertex positions
through a locally differentiable map.
We exploit this distinction by preserving
discrete edge selections and connectivity
updates during forward simplification,
while differentiating the resulting geometry
in backward.

Let $\mathcal{M}^{(0)}=\mathcal{M}_\theta$,
and let $\sigma(V^{(0)},F^{(0)})$ denote
the discrete collapse history, recording
the collapsed edge sets $\{\mathcal{E}_r\}_{r=1}^R$,
pass boundaries, surviving and removed
endpoint indices, and the placement branch of each collapse.
With $\sigma$ fixed, vertex placement,
quadric addition, and pass-wise quadric
recomputation compose into a locally
differentiable \emph{collapse map}
$\mathcal{C}_\sigma:
V^{(0)}\mapsto V^{(R)}$.
We can therefore express the forward
simplification operator as
\begin{equation}
    \mathcal{S}_K(\mathcal{M}^{(0)})
    =
    \mathcal{M}^{(R)}
    =
    \left(
        \mathcal{C}_\sigma(V^{(0)}),
        F^{(R)}
    \right).
    \label{eq:qem_decomposition}
\end{equation}

Substituting Eq.~\eqref{eq:qem_decomposition}
into Eq.~\eqref{eq:optimization_objective},
the loss becomes
$\mathcal{L}(\mathcal{C}_\sigma(V_\theta),F^{(R)})$.
Where $\sigma$ remains unchanged,
$F^{(R)}$ is fixed, and differentiating
the loss through the output vertex
positions yields
\begin{equation}
\begin{aligned}
\label{eq:fixed_history_gradient}
\nabla_\theta
\mathcal{L}(\mathcal{S}_K(\mathcal{M}_\theta))
&=
\left(
    \frac{\partial V_\theta}{\partial\theta}
\right)^{\!\top}
\left(
    \frac{\partial\mathcal{C}_\sigma(V_\theta)}
         {\partial V_\theta}
\right)^{\!\top}
\nabla_{V^{(R)}}\mathcal{L}.
\end{aligned}
\end{equation}
The collapse-map Jacobian
$\frac{\partial\mathcal{C}_\sigma(V_\theta)}
{\partial V_\theta}$
captures dependencies across successive
collapses, allowing gradients to reach
input vertices removed during simplification.

\subsection{Factorized Parallel Differentiation}
\label{sec:method:parallel}

Efficient evaluation of the VJP
$\left(
\frac{\partial\mathcal C_\sigma(V_\theta)}
{\partial V_\theta}
\right)^\top
\nabla_{V^{(R)}}\mathcal L$
in Eq.~\eqref{eq:fixed_history_gradient}
requires handling geometric dependencies
in the collapse-map Jacobian, both within
each round and across successive rounds.
Within a round, the independent regions
$\{\mathcal{N}_1(e)\}_{e\in\mathcal{E}_r}$
are face-disjoint but may share neighboring vertices,
inducing overlapping dependencies
through face quadrics.
Across rounds, collapses depend
on previously updated vertex positions
and accumulated vertex quadrics.
We address these dependencies by treating
vertex quadrics as differentiable
intermediate variables, separating each
collapse from its underlying face geometry.
This allows gradients for multiple collapses
within a round to be computed in parallel.
Across rounds, vertex quadric gradients
are propagated through additive quadric
merges along a reduction forest, avoiding
redundant differentiation.

Let $P$ denote the number of simplification
passes, $V_p$ the output vertex positions
of pass $p$ with $V_0=V^{(0)}=V_\theta$,
and $\mathbf Q$ the collection of vertex quadrics $Q_i$.
At the beginning of pass $p$, the vertex quadrics
$\mathbf Q_{p-1}$ are recomputed from
the input geometry $V_{p-1}$.
For pass $p$ with $R_p$ collapse rounds,
we use local round indices and omit
the pass index.
Let $X^{(r)}=(V^{(r)},\mathbf Q^{(r)})$
denote the state after round $r$, initialized
as $X^{(0)}=(V_{p-1},\mathbf Q_{p-1})$, and let
$J_r=\partial X^{(r)}/\partial X^{(r-1)}$
be its round-wise Jacobian.
The Jacobian of pass $p$ factorizes as
\begin{equation}
    D_p :=
    \frac{\partial V_p}{\partial V_{p-1}}
    =
    \begin{bmatrix}I&0\end{bmatrix}
    J_{R_p}\cdots J_1
    \begin{bmatrix}
        I\\
        \dfrac{\partial\mathbf Q_{p-1}}
              {\partial V_{p-1}}
    \end{bmatrix},
    \label{eq:collapse_jacobian_factorization}
\end{equation}
where the leftmost factor selects the
output vertex positions, and the rightmost
factor differentiates the pass initialization,
including quadric recomputation.
The full collapse-map Jacobian is then
$\frac{\partial\mathcal C_\sigma(V_\theta)}
{\partial V_\theta}=D_P\cdots D_1$.

\paragraph{Parallel Jacobian Structure.} %
For $e=(i,j)\in\mathcal E_r$ within pass $p$,
let $J_e^{(r)}$ denote the Jacobian of
the collapse update
$(v_i^{(r-1)},Q_i^{(r-1)},
 v_j^{(r-1)},Q_j^{(r-1)})
\mapsto (v_e^{(r)},Q_e^{(r)}).
$
Since the face-derived geometric contributions
to each collapse are accumulated in its
endpoint quadrics, treating these quadrics
as differentiable intermediate variables
confines $J_e^{(r)}$ to its two endpoint states.
As the collapsed edges have face-disjoint
regions $\mathcal N_1(e)$ and hence disjoint
endpoints, the round-wise Jacobian separates
into independent collapse-local blocks:
\begin{equation}
    J_r =
    \left(
        \bigoplus_{e\in\mathcal E_r}
        J_e^{(r)}
    \right)\oplus I,
    \label{eq:round_jacobian}
\end{equation}
up to permutations, where $\oplus$
denotes the matrix direct sum and $I$
is the identity block for unchanged
vertex states.
Consequently, the round-wise VJP decomposes
into independent collapse-local VJPs,
evaluated in parallel without conflicting
writes to endpoint gradients.

\paragraph{Quadric Reduction-Forest Backpropagation.}
While the round-wise Jacobians enable
parallel local differentiation,
accumulated vertex quadrics retain dependencies
across successive collapses.
Since vertex quadrics are merged additively,
the recorded merge history forms a
reduction forest whose leaves are the initial
vertex quadrics.
We propagate quadric gradients backward through
this forest, deferring face-level differentiation
until its leaves are reached.
This avoids repeatedly differentiating the
contributing face geometry at each collapse.

Let $g^{(r)}=\nabla_{X^{(r)}}\mathcal L$
denote the gradient with respect to
the state after round $r$.
Since the loss depends on the terminal state
$X^{(R_p)}$ only through $V_p$,
we initialize
$g^{(R_p)}=(\nabla_{V_p}\mathcal L,0)$
and apply
$g^{(r-1)}=J_r^\top g^{(r)}$
in reverse round order.
Placement derivatives generate quadric gradients
that are propagated through earlier merges.
For $e=(i,j)\in\mathcal E_r$,
let $g_{v_e}^{(r)}$ and $g_{Q_e}^{(r)}$
denote the incoming gradients at its
merged position and quadric.
Since $Q_e^{(r)}=Q_i^{(r-1)}+Q_j^{(r-1)}$,
both input vertex quadrics receive identical gradients:
\begin{equation}
    g_{Q_i}^{(r-1)}
    =g_{Q_j}^{(r-1)}
    =
    g_{Q_e}^{(r)}
    +
    \left(
        \frac{\partial v_e^{(r)}}
             {\partial Q_e^{(r)}}
    \right)^\top
    g_{v_e}^{(r)}.
\label{eq:quadric_merge_gradient}
\end{equation}
The first term carries quadric gradients from
subsequent collapses, while the second adds
the contribution from the current vertex placement.

Propagating gradients through the
merge history yields $g^{(0)}$,
the gradient with respect to $V_{p-1}$
and $\mathbf Q_{p-1}$ treated as
independent state inputs.
We then differentiate the quadric recomputation
for this pass to obtain the full gradient
with respect to the input geometry.
Since $Q_i^{(0)}=\sum_{f\ni i}K_f$,
\begin{equation}
\begin{aligned}
    g_{K_f}
    &=
    \sum_{i\in f}g_{Q_i}^{(0)},\\
    \nabla_{v_i^{(0)}}\mathcal L
    &=
    g_{v_i}^{(0)}
    +
    \sum_{f\ni i}
    \left(
        \frac{\partial K_f}
             {\partial v_i^{(0)}}
    \right)^\top
    g_{K_f}.
\end{aligned}
\label{eq:initial_vertex_gradient}
\end{equation}
This combines direct position gradients
with deferred quadric-induced gradients,
yielding the pass-wise VJP
$D_p^\top\nabla_{V_p}\mathcal L
=\nabla_{V_{p-1}}\mathcal L$.
Applying these pass-wise VJPs in reverse
pass order yields the full collapse-map VJP
in Eq.~\eqref{eq:fixed_history_gradient},
including differentiation through quadric
recomputation at each pass boundary.

Within each pass, each collapse requires
a constant-size local backward update,
and each initial face is differentiated only once.
These backward updates therefore require
$O(\sum_{r=1}^{R_p}|\mathcal E_r|+|F_{p-1}|)$
work, where $F_{p-1}$ is the input face set
of pass $p$. Even when retaining only 0.5\% of the input
vertices, simplification takes about four
passes on average.
Our CUDA implementation performs these
updates without constructing per-collapse
PyTorch autograd graphs~\cite{paszke2019pytorch}.
The placement derivatives used to propagate
position gradients are provided in
Appendix~B.

\subsection{Integration into Gradient-Based Optimization}
\label{sec:method:scheduling}

Applying aggressive simplification before
the underlying geometry is sufficiently optimized,
or abruptly imposing the target vertex budget,
can cause substantial geometric changes
and destabilize optimization.
We therefore introduce a progressive vertex-budget
schedule for integrating simplification into
gradient-based geometry optimization.
Let $t$ index optimization iterations
and $N_t=|V_\theta|=|V^{(0)}|$
denote the input vertex count at iteration $t$.
With warmup duration $t_w$ and budget-reduction
duration $t_r$, we define
\begin{equation}
    s_t =
    \begin{cases}
        0, & t<t_w,\\
        \dfrac{t-t_w}{t_r},
        & t_w\leq t<t_w+t_r,\\
        1, & t\geq t_w+t_r.
    \end{cases}
    \label{eq:budget_schedule_factor}
\end{equation}
The vertex budget at iteration $t$ is
\begin{equation}
    K_t =
    \operatorname{round}
    \left(N_t^{1-s_t}K^{s_t}\right),
    \label{eq:effective_budget}
\end{equation}
where $\operatorname{round}(\cdot)$ denotes rounding to the
nearest integer.
After warmup, the vertex budget is
interpolated in log space from $N_t$ to $K$,
followed by optimization at the target budget.
At each iteration, we apply
$\mathcal S_{K_t}$ to the current mesh.

Although the collapse history may change
between optimization iterations, we empirically
demonstrate that differentiating the collapse
map under each realized history supports
effective optimization (Sec.~\ref{sec:exp:analysis}).

\begin{figure*}[t]
    \centering
    \includegraphics[width=\textwidth]{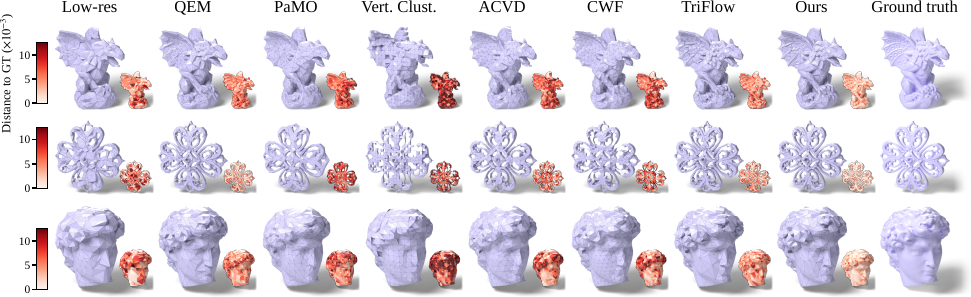}
    \vspace{-1.5em}
    \caption{Qualitative comparison at $K=1000$ corresponding to Table~\ref{tab:recon_main}.
    }
    \label{fig:teaser}
\end{figure*}

\section{Experiments}
\label{sec:experiments}

We evaluate \methodname{} on low-poly mesh
reconstruction and inverse rendering.
In both tasks, we follow the progressive
vertex-budget schedule in Sec.~\ref{sec:method:scheduling},
with warmup duration $t_w=0.2T$ and
budget-reduction duration $t_r=0.45T$,
where $T$ is the total number of optimization iterations.
All experiments are conducted on a single
RTX~5090. For detailed experimental settings,
see Appendix~A.

\subsection{Low-Poly Mesh Reconstruction}
\label{subsec:low-poly}
\paragraph{Experimental Setup.}
We use 79 shapes from Myles et al.~\cite{myles2014robust},
excluding non-watertight and very thin
shapes~\cite{shen2023flexible}.
Our base reconstruction pipeline optimizes
FlexiCubes~\cite{shen2023flexible}
at grid resolution $n=128$ for 5,000 iterations
on shapes normalized to fit within $[-0.9,0.9]^3$,
using multi-view silhouette, depth, and normal supervision.
Baselines include low-resolution FlexiCubes
(Low-res), using the finest per-shape grid
with at most $K$ vertices, and representative
mesh simplification methods applied to the
base reconstructions: QEM~\cite{garland1997surface}
(MeshLab~\cite{cignoni2008meshlab}),
PaMO~\cite{oh2025pamo}, and
vertex clustering~\cite{rossignac1993multiresolution},
along with the remeshing method
ACVD~\cite{valette2004approximated}.
We also compare against CWF~\cite{xu2024cwf},
which optimizes simplified geometry, and
TriFlow~\cite{li2026triflow},
a learning-based method.
Unlike post-processing, we integrate
\methodname{} into the base pipeline
to optimize through the simplified mesh.

We evaluate the largest connected component
to exclude disconnected interior geometry
unseen by image-based supervision~\cite{binninger2025tetweave}.
We report squared Chamfer distance (CD)
and F1 ($10^{-3}$ threshold)
using $10^6$ surface samples,
along with normal consistency (NC),
self-intersection rate (SI), and sliver-triangle rate (SR).
CD is scaled by $10^5$, and SR is the percentage
of triangles with minimum angle below $1^\circ$.

\begin{figure}[t]
    \centering
    \includegraphics[width=0.98\columnwidth]{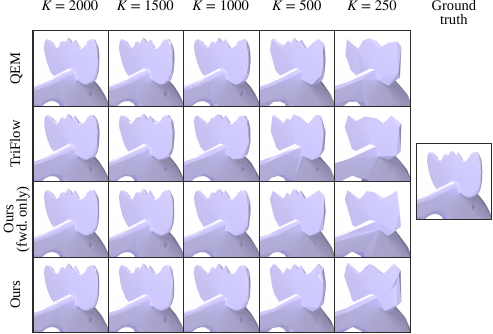}
    \caption{Qualitative comparison across vertex budgets (Table~\ref{tab:recon_budget}).}
    \label{fig:recon_budget}
\end{figure}

\paragraph{Comparison with Low-Poly Modeling Methods.}
Table~\ref{tab:recon_main} and Fig.~\ref{fig:teaser}
compare low-poly modeling methods at $K=1000$.
Our method outperforms all baselines in CD, F1, and NC, while maintaining SI and SR
comparable to those of QEM.
These results demonstrate the effectiveness
of optimizing through simplification
over both direct low-resolution reconstruction
and post-processing simplification. %
We further evaluate reconstruction across
vertex budgets $K\in\{2000,1500,1000,500,250\}$
at $n=128$, keeping the remaining settings unchanged.
To distinguish the contributions of forward
simplification and differentiation,
we also evaluate our forward operator
as post-processing (Ours, fwd. only).
As shown in Table~\ref{tab:recon_budget}
and Fig.~\ref{fig:recon_budget},
differentiating through simplification
consistently improves reconstruction fidelity
over the forward-only variant.
At larger budgets, most of the improvement
over QEM comes from the forward simplifier,
since simplification-induced errors are small
and leave little for gradients to correct.
Under tighter budgets, these errors grow
and gradients compensate for more of them,
so the absolute CD gain from differentiation increases
from $0.05$ at $K=2000$ to $1.34$ at $K=250$.
\begin{table}[t]
  \centering

\caption{Low-poly reconstruction with $n=128$, $K=1000$.
Reference denotes the unsimplified base reconstruction.
CD$_{\mathrm{mean}}$ and CD$_{\mathrm{med}}$
denote mean and median CD ($\times 10^{-5}$).}

  \label{tab:recon_main}

\footnotesize
\setlength{\tabcolsep}{1.5pt}
\renewcommand{\arraystretch}{1.1}
  \begin{tabular}{@{}lrrrrrrr@{}}
  \toprule
  Method
  & CD$_{\mathrm{mean}}\downarrow$
  & CD$_{\mathrm{med}}\downarrow$
  & F1$\uparrow$
  & NC$\uparrow$
  & SI(\%)$\downarrow$
  & SR(\%)$\downarrow$
  & $|V|$ \\
  \midrule
  Reference
  & 0.47 & 0.39 & 0.45 & 0.99
  & 0.70 & 0.65 & 42,195 \\
  \midrule
  Low-res~\cite{shen2023flexible}
  & 338.22 & 9.02 & 0.20 & 0.93
  & 6.41 & 0.93 & 924 \\
  QEM~\cite{garland1997surface}
  & 1.92 & 1.11 & 0.21 & 0.96
  & 0.56 & 0.02 & 989 \\
  PaMO~\cite{oh2025pamo}
  & 4.96 & 2.90 & 0.12 & 0.95
  & \textbf{0.00} & \textbf{0.00} & 988 \\
  Vert. Clust.~\cite{rossignac1993multiresolution}
  & 42.99 & 17.76 & 0.06 & 0.89
  & 10.02 & 0.01 & 995 \\
  ACVD~\cite{valette2004approximated}
  & 4.35 & 2.62 & 0.17 & 0.95
  & 0.64 & 0.14 & 1,047 \\
  CWF~\cite{xu2024cwf}
  & 7.10 & 5.15 & 0.11 & 0.94
  & 0.02 & 0.02 & 1,000 \\
  TriFlow~\cite{li2026triflow}
  & 2.53 & 1.20 & 0.19 & 0.96
  & 2.67 & 0.37 & 989 \\
  \midrule
  Ours
  & \textbf{1.26} & \textbf{0.76}
  & \textbf{0.26} & \textbf{0.97}
  & 0.62 & 0.05 & 1,000 \\
  \bottomrule
  \end{tabular}
  \end{table}

\begin{table}[t]
  \centering
\caption{Reconstruction across vertex budgets
at $n=128$ (CD $\times 10^{-5}$).
Ours (fwd. only) applies our simplifier
as post-processing.
$\Delta_{\mathrm{grad}}$ denotes the absolute
gain over this variant, with positive values
indicating improvement.}
  \label{tab:recon_budget}

  \footnotesize
  \setlength{\tabcolsep}{2pt}

  \begin{tabular}{@{}clccccccc@{}}
  \toprule
  $K$ & Metric
  & QEM & ACVD & CWF & TriFlow
  & \shortstack{Ours\\(fwd. only)}
  & Ours & $\Delta_{\mathrm{grad}}\uparrow$ \\
  \midrule
  2000 & CD$\downarrow$
  & 0.90 & 1.96 & 3.25 & 0.95 & 0.75
  & \textbf{0.70} & $+0.05$ \\
       & F1$\uparrow$
  & 0.30 & 0.25 & 0.18 & 0.27 & 0.31
  & \textbf{0.34} & $+0.03$ \\
\midrule
  1500 & CD$\downarrow$
  & 1.18 & 2.69 & 4.42 & 1.29 & 0.95
  & \textbf{0.87} & $+0.08$ \\
       & F1$\uparrow$
  & 0.26 & 0.22 & 0.15 & 0.23 & 0.28
  & \textbf{0.31} & $+0.03$ \\
  \midrule
  1000 & CD$\downarrow$
  & 1.92 & 4.35 & 7.10 & 2.53 & 1.53
  & \textbf{1.26} & $+0.27$ \\
       & F1$\uparrow$
  & 0.21 & 0.17 & 0.11 & 0.19 & 0.23
  & \textbf{0.26} & $+0.03$ \\
  \midrule
  500 & CD$\downarrow$
  & 5.10 & 10.72 & 26.41 & 6.78 & 3.97
  & \textbf{2.89} & $+1.08$ \\
      & F1$\uparrow$
  & 0.14 & 0.10 & 0.06 & 0.12 & 0.16
  & \textbf{0.18} & $+0.03$ \\
  \midrule
  250 & CD$\downarrow$
  & 14.88 & 36.59 & 111.05 & 22.87 & 12.10
  & \textbf{10.76} & $+1.34$ \\
      & F1$\uparrow$
  & 0.09 & 0.06 & 0.04 & 0.07 & 0.10
  & \textbf{0.13} & $+0.03$ \\
  \bottomrule
  \end{tabular}
\end{table}

\paragraph{Integration across Geometry Optimization Pipelines.}
We evaluate \methodname beyond FlexiCubes
by replacing FlexiCubes with
DiffMC~\cite{wei2025neumanifold},
integrating it into the separate tetrahedral
reconstruction pipeline of DMTet~\cite{shen2021deep},
and applying it to Large Steps~\cite{nicolet2021large},
which optimizes explicit meshes through
differential coordinates without isosurface extraction.
All three use the same multi-view supervision
as the base reconstruction pipeline.
We use $K = 1000$ ($n = 128$ for DMTet and DiffMC) and the 42 genus-zero shapes for Large Steps.
Table~\ref{tab:recon_frontends}
shows that \methodname outperforms all baselines
in both CD and F1 across the three pipelines.
These results demonstrate that optimizing
through simplification consistently improves
low-poly reconstruction across different
gradient-based geometry optimization pipelines,
extending beyond isosurface extraction
to explicit mesh optimization.

\begin{table}[t]
  \centering
  \caption{Low-poly reconstruction across
  geometry optimization pipelines at $K=1000$
  (CD $\times 10^{-5}$).
  DMTet and DiffMC use $n=128$, while
  Large Steps uses a fixed-topology icosphere.}
  \label{tab:recon_frontends}

  \footnotesize
  \setlength{\tabcolsep}{2pt}
  \renewcommand{\arraystretch}{1.1}

  \resizebox{\columnwidth}{!}{%
  \begin{tabular}{@{}llcccccc@{}}
    \toprule
    Frontend & Metric
    & QEM & ACVD & CWF & TriFlow
    & \shortstack{Ours\\(fwd. only)} & Ours \\
    \midrule
    DMTet~\cite{shen2021deep} & CD$\downarrow$
    & 1.82 & 4.42 & 7.01 & 2.85
    & 1.56 & \textbf{1.23} \\
    & F1$\uparrow$
    & 0.21 & 0.16 & 0.12 & 0.19
    & 0.23 & \textbf{0.26} \\
    \midrule
    DiffMC~\cite{wei2025neumanifold} & CD$\downarrow$
    & 2.07 & 4.27 & 7.46 & 1.98
    & 1.44 & \textbf{1.03} \\
    & F1$\uparrow$
    & 0.21 & 0.17 & 0.11 & 0.19
    & 0.23 & \textbf{0.29} \\
    \midrule
    Large Steps~\cite{nicolet2021large} & CD$\downarrow$
    & 7.35 & 11.05 & 9.41 & 7.31
    & 6.81 & \textbf{6.47} \\
    & F1$\uparrow$
    & 0.14 & 0.10 & 0.09 & 0.13
    & 0.15 & \textbf{0.21} \\
    \bottomrule
  \end{tabular}%
  }
\end{table}

\subsection{Inverse Rendering}
\label{sec:exp:inverse_rendering}

\paragraph{Experimental Setup.}
We use nvdiffrec~\cite{munkberg2022extracting}
as our base inverse-rendering pipeline,
with FlexiCubes as the geometry frontend.
We evaluate eight NeRF-synthetic
scenes~\cite{mildenhall2020nerf}.
Stage 1 optimizes geometry, materials, and lighting,
followed by UV parameterization and texture baking.
Stage 2 jointly refines vertex positions
and UV textures with fixed connectivity
for 5,000 iterations.
We integrate \methodname into Stage 1 geometry
optimization while retaining the same Stage 2.
At $K=2000$, we compare against
low-resolution FlexiCubes (Low-res),
QEM~\cite{garland1997surface},
and our forward-only variant.
Low-res directly optimizes a coarse grid,
whereas QEM and Ours (fwd. only) simplify
the geometry after Stage 1 optimization
but before texture baking.
All three undergo the same Stage 2
refinement as our method.
We additionally compare with texture-aware
QEM (TQEM)~\cite{garland1998simplifying}
and appearance-driven simplification
(ADS)~\cite{hasselgren2021appearance},
which operate on the final textured reference.
ADS further refines the TQEM-initialized
mesh to match the reference.
We report PSNR, SSIM, and LPIPS
on held-out views.

\begin{figure}[t]
    \centering
    \includegraphics[width=\columnwidth]{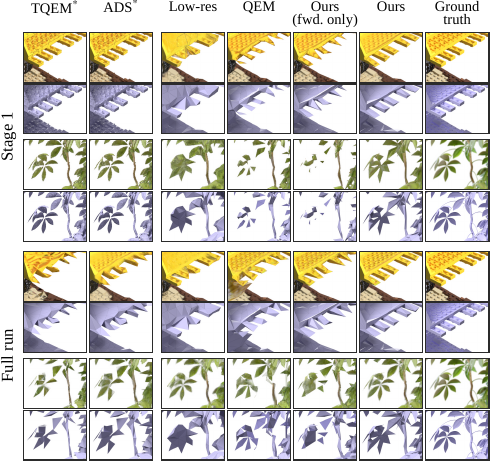}
\caption{Qualitative inverse rendering results (Table~\ref{tab:inverse_rendering}).
Stage 1 entries for methods marked $^*$
are inherited from Reference.}
    \label{fig:irender}
\end{figure}

\paragraph{Rendering Fidelity after Simplification.}
As shown in Table~\ref{tab:inverse_rendering},
after Stage 1, \methodname is only 1.32 dB
below the dense reference while retaining
just 3.9\% of its vertices.
As illustrated in Fig.~\ref{fig:irender},
our method better preserves fine structures
both after Stage 1 and after the full run.
It outperforms both post-processing QEM and
our forward-only variant, demonstrating
that the gains extend beyond improvements
in the forward simplifier alone.
These advantages persist after the shared Stage 2 refinement.
Our method also outperforms TQEM and ADS
across all rendering metrics, indicating that
appearance-aware post-processing does not
match the fidelity achieved by optimizing
upstream geometry through differentiable
simplification.
These results demonstrate that
optimizing through simplification enables
upstream geometry to compensate for
simplification-induced errors, improving the final textured mesh.

\begin{table}[t]
  \centering
\caption{Novel-view synthesis on eight NeRF-synthetic
scenes ($K=2000$).
Full run includes Stage 2 refinement.
Methods marked $^*$ simplify the final textured
reference; their Stage 1 results are inherited from Reference and omitted.}
  \label{tab:inverse_rendering}
  \scriptsize
  \setlength{\tabcolsep}{1.5pt}
  \renewcommand{\arraystretch}{1.15}
  \resizebox{\columnwidth}{!}{%
  \begin{tabular}{lcccccccc}
  \toprule
  & \multicolumn{4}{c}{Stage 1}
  & \multicolumn{4}{c}{Full run} \\
  \cmidrule(lr){2-5}
  \cmidrule(lr){6-9}
  Method
  & PSNR$\uparrow$ & SSIM$\uparrow$ & LPIPS$\downarrow$ & $|V|$
  & PSNR$\uparrow$ & SSIM$\uparrow$ & LPIPS$\downarrow$ & $|V|$ \\
  \midrule
  Reference
  & 26.88 & 0.907 & 0.114 & 51,512
  & 29.02 & 0.937 & 0.082 & 51,512 \\
  \midrule
  TQEM$^{*}$~\cite{garland1998simplifying}
  & -- & -- & -- & --
  & 22.47 & 0.853 & 0.139 & 2,000 \\
  ADS$^{*}$~\cite{hasselgren2021appearance}
  & -- & -- & -- & --
  & 25.71 & 0.907 & 0.110 & 2,000 \\
  Low-res
  & 23.73 & 0.880 & 0.132 & 1,883
  & 25.56 & 0.907 & 0.106 & 1,883 \\
  QEM~\cite{garland1997surface}
  & 23.73 & 0.885 & 0.123 & 2,000
  & 25.43 & 0.907 & 0.100 & 2,000 \\
  Ours (fwd. only)
  & 24.18 & 0.889 & 0.120 & 2,000
  & 26.03 & 0.912 & 0.095 & 2,000 \\
  \midrule
  Ours
  & \textbf{25.56} & \textbf{0.906} & \textbf{0.109} & 2,000
  & \textbf{27.29} & \textbf{0.925} & \textbf{0.085} & 2,000 \\
  \bottomrule
  \end{tabular}%
  }
\end{table}

\subsection{Method Analysis}
\label{sec:exp:analysis}

\begin{table}[t]
\centering
\caption{Effect of backward differentiation on
79 shapes.
CD ($\times 10^{-5}$) and F1 evaluate the final
simplified meshes.
Raw and Deployed denote rendering objectives
evaluated on the unsimplified and $K=500$
meshes, respectively.
We report mean relative objective changes
($\Delta L/L$, \%) over budget reduction
(Ramp, iterations 1000--3250),
target-budget optimization (Hold, 3250--5000),
and the entire interval (Overall, 1000--5000).}
\label{tab:exp:causal}
\scriptsize
\setlength{\tabcolsep}{2.5pt}
\renewcommand{\arraystretch}{1.1}
\begin{tabular}{lcccccc}
\toprule
& & & Raw $\Delta L/L$
& \multicolumn{3}{c}{Deployed $\Delta L/L$} \\
\cmidrule(lr){5-7}
Variant & CD$\downarrow$ & F1$\uparrow$
& Hold & Ramp & Hold & Overall \\
\midrule
Forward-only & 3.971 & 0.156
& $-3.4$ & $+0.6$ & $+0.4$ & $+1.2$ \\
Survivor grad & 3.652 & 0.152
& $+6.4$ & $-5.5$ & $-2.4$ & $-6.9$ \\
Cluster grad & 3.533 & 0.177
& $+19.3$ & $\mathbf{-29.7}$ & $-2.1$ & $-30.2$ \\
W/o quadric grad & 3.941 & 0.141
& $+3.4$ & $+20.6$ & $+0.5$ & $+21.6$ \\
\midrule
Ours & \textbf{2.891} & \textbf{0.182}
& $+12.3$ & $-26.3$ & $\mathbf{-9.7}$
& $\mathbf{-33.2}$ \\
\bottomrule
\end{tabular}
\end{table}

\paragraph{Effectiveness of Simplification Gradients.}
We evaluate the effect of our gradients using the base
reconstruction pipeline in Sec.~\ref{subsec:low-poly}
at $K=500$.
We compare against \textit{Forward-only} post-processing
and three in-loop variants sharing our forward simplifier:
\textit{Survivor grad} propagates each output gradient
only to its surviving input vertex,
\textit{Cluster grad} distributes it equally among
all input vertices merged into that output vertex, and
\textit{W/o quadric grad} treats quadrics as constants
while retaining direct vertex-position gradients.
Following our progressive vertex-budget schedule,
we measure optimization progress over three intervals:
Ramp (iterations 1000--3250, budget reduction),
Hold (3250--5000, fixed target budget),
and Overall (1000--5000).
At iterations 1000, 3250, and 5000, we evaluate
rendering losses against ground-truth renderings
using the unsimplified mesh (Raw) and its
simplified version at $K=500$ (Deployed),
regardless of the current vertex budget.
We report relative changes
$\Delta L/L=(L_{\mathrm{end}}-L_{\mathrm{start}})/
L_{\mathrm{start}}$.

Table~\ref{tab:exp:causal} shows that all in-loop
variants improve CD over Forward-only, while
Cluster grad and Ours also improve F1,
suggesting that optimizing through simplification
can improve geometric fidelity.
Cluster grad makes substantial progress during
Ramp but largely stalls during Hold.
Survivor grad yields limited gains, while removing
quadric gradients worsens the deployed objective.
In contrast, our method continues to reduce
the deployed objective, achieving the greatest
overall improvement and best geometric fidelity.
These results highlight the importance of
differentiating the geometric dependencies
of simplification rather than relying on
surrogate gradient propagation.
Moreover, gradient-based optimization remains effective
even as collapse histories are reselected
at every iteration.

The raw and deployed objectives further expose
the optimization--deployment mismatch.
Forward-only reduces the raw objective without
improving the deployed objective, whereas our method
reduces the deployed objective even as its raw
objective increases.
This divergence shows that optimizing unsimplified
geometry does not necessarily improve its
budget-constrained output.
Optimizing through simplification instead enables
upstream geometry to compensate for errors
introduced by the final complexity reduction.

  \begin{table}[t]
  \centering
\caption{Operator runtime on Gaussian random-field
meshes at $K=1000$, measured in ms.
Speedups are relative to Ours, and Launch$\times$
denotes the backward CUDA kernel-launch ratio
(Autograd/Ours).
QEM runs on CPU, while the others run on GPU.}
  \label{tab:operator_runtime}
  \scriptsize
  \setlength{\tabcolsep}{2pt}
  \renewcommand{\arraystretch}{1.1}
  \resizebox{\columnwidth}{!}{%
  \begin{tabular}{rrrrrrrr}
  \toprule
  & \multicolumn{3}{c}{Forward}
  & \multicolumn{4}{c}{Backward} \\
  \cmidrule(lr){2-4}
  \cmidrule(lr){5-8}
  $|V_{\mathrm{in}}|$
  & Ours & QEM & Speedup
  & Ours & Autograd & Speedup & Launch$\times$ \\
  \midrule
  50,145
  & 8.65 & 530.32 & 61.3$\times$
  & 1.99 & 74.26 & 37.3$\times$ & 81.5$\times$ \\
  502,141
  & 36.37 & 8,198.28 & 225.4$\times$
  & 2.34 & 144.53 & 61.8$\times$ & 107.0$\times$ \\
  5,045,762
  & 322.21 & 102,830.06 & 319.1$\times$
  & 6.75 & 509.20 & 75.4$\times$ & 122.5$\times$ \\
  \bottomrule
  \end{tabular}%
  }
  \end{table}

\paragraph{Computational Efficiency.}
We compare \methodname against QEM and a
PyTorch autograd replay on Gaussian random-field meshes.
As shown in Table~\ref{tab:operator_runtime},
our forward pass is 61.3--319.1$\times$ faster
than QEM, demonstrating the efficiency
of parallel GPU execution.
Under identical collapse histories,
the order-of-magnitude backward speedups
and fewer CUDA kernel launches reflect
our factorized backward, which enables parallel
collapse-local VJPs and avoids redundant
face-level differentiation across rounds
through the quadric reduction forest.
For low-poly reconstruction in Table~\ref{tab:recon_main},
\methodname incurs $1.25\times$ the runtime
of its base pipeline (205 versus 164 seconds).
For inverse rendering in Table~\ref{tab:inverse_rendering},
its runtime remains comparable to the base pipeline (3,262 versus 3,255 seconds).
Peak GPU memory increases by 11 and 59 MiB, respectively.

\section{Conclusion}
\label{sec:conclusion}

We present \methodname, integrating QEM-based mesh
simplification into gradient-based geometry
optimization as a differentiable layer.
By differentiating the collapse map with the
recorded collapse history fixed, our method
enables upstream optimization to compensate
for simplification-induced errors.
Combining our improved forward simplifier with
a factorized backward scheme, \methodname outperforms
the evaluated post-processing simplification baselines
in low-poly reconstruction and inverse rendering,
with modest runtime and memory overhead.
We further show that gradient propagation provides
increasing gains under tighter vertex budgets,
where simplification introduces larger geometric errors.
These improvements extend across diverse geometry
optimization frontends, demonstrating the effectiveness
of optimizing through \methodname rather than
applying simplification as post-processing.

\section*{Acknowledgments}
We thank Xinyue Wei for insightful discussions and valuable feedback that helped shape and refine this work.

{
    \small
    \bibliographystyle{ieeenat_fullname}
    \bibliography{main}
}
\clearpage
\onecolumn

\begin{center}
  {\Large\textbf{\thetitle}}\\[0.5em]
  {\Large Supplementary Material}
\end{center}
\vspace{1.0em}

\appendix
\setcounter{table}{0}
\setcounter{figure}{0}
\setcounter{equation}{0}
\setcounter{footnote}{0}
\renewcommand{\thetable}{S\arabic{table}}
\renewcommand{\thefigure}{S\arabic{figure}}
\renewcommand{\theequation}{S\arabic{equation}}
\providecommand{\theHtable}{}
\providecommand{\theHfigure}{}
\providecommand{\theHequation}{}
\renewcommand{\theHtable}{S\arabic{table}}
\renewcommand{\theHfigure}{S\arabic{figure}}
\renewcommand{\theHequation}{S\arabic{equation}}

\startcontents[supp]
\titlecontents{lsection}[0em]{\addvspace{0.7em}\bfseries}
  {\thecontentslabel\hspace*{1em}}{}{\titlerule*[0.6pc]{.}\contentspage}
\titlecontents{lsubsection}[1.5em]{\addvspace{0.2em}}
  {\thecontentslabel\hspace*{1em}}{}{\titlerule*[0.6pc]{.}\contentspage}
\titlecontents{lsubsubsection}[3.0em]{}
  {\thecontentslabel\hspace*{1em}}{}{\titlerule*[0.6pc]{.}\contentspage}
\noindent{\large\bfseries Contents}\par\vspace{0.3em}
{\hypersetup{linkcolor=black}%
 \printcontents[supp]{l}{1}{\setcounter{tocdepth}{3}}}
\vspace{1.0em}

\FloatBarrier                     %

\section{Experimental Details}
\label{app:experimental}
\label{app:implementation}

\subsection{Low-Poly Mesh Reconstruction}
\label{app:exp:low-poly}
For the low-poly mesh reconstruction experiments in
Sec.~\ref{subsec:low-poly}, we build on the official
FlexiCubes implementation, optimizing a randomly
initialized SDF at grid resolution $n=128$ for
$T=5000$ iterations using multi-view silhouette,
depth, and normal supervision without ground-truth
SDF.
Our reconstruction objective is
\begin{equation}
\begin{aligned}
\mathcal{L}
={}&\lambda_{\mathrm{mask}}\mathcal{L}_{\mathrm{mask}}
+\lambda_{\mathrm{depth}}\mathcal{L}_{\mathrm{depth}}
+\lambda_{\mathrm{normal}}\mathcal{L}_{\mathrm{normal}}
+\lambda_{\mathrm{dev}}\mathcal{L}_{\mathrm{dev}}
+\lambda_{\mathrm{sign}}\mathcal{L}_{\mathrm{sign}},
\end{aligned}
\label{eq:app:recon_objective}
\end{equation}
where $\mathcal{L}_{\mathrm{dev}}$ and
$\mathcal{L}_{\mathrm{sign}}$ are the FlexiCubes
deformation and sign regularizers.
We set $\lambda_{\mathrm{mask}}=5$,
$\lambda_{\mathrm{depth}}=400$,
$\lambda_{\mathrm{normal}}=1$, and
$\lambda_{\mathrm{dev}}=1$.
Following the official FlexiCubes implementation,
we linearly decay $\lambda_{\mathrm{sign}}$
from $0.2$ to $0.01$ over the first $0.25T$ iterations.
The other reconstruction pipelines in
Table~\ref{tab:recon_frontends} use the same
multi-view supervision scheme with frontend-specific settings.
DiffMC follows our FlexiCubes reconstruction setup without
$\mathcal{L}_{\mathrm{dev}}$.
DMTet also omits $\mathcal{L}_{\mathrm{dev}}$,
but uses $\lambda_{\mathrm{mask}}=10$,
$\lambda_{\mathrm{depth}}=250$, and
$\lambda_{\mathrm{normal}}=1$, while retaining
the same sign regularization schedule.
Large Steps uses the same rendering loss weights
as DMTet, without deformation or sign regularization.
Within each frontend, our method and the
post-processing baselines share the same
base reconstruction settings.

\subsection{Inverse Rendering}
\label{appendix:inverse_rendering_setup}
For inverse rendering in Sec.~\ref{sec:exp:inverse_rendering},
we follow the official nvdiffrec configurations
for the eight NeRF-synthetic scenes, using FlexiCubes
at $n=128$. Both stages run for 5,000 iterations
with the original inverse-rendering objective,
\begin{equation}
    \mathcal{L}
    =\mathcal{L}_{\mathrm{image}}
    +\mathcal{L}_{\mathrm{mask}}
    +\lambda\mathcal{L}_{\mathrm{reg}},
\end{equation}
where the regularization terms follow the official
configurations.
For the post-processing ADS baseline, which requires
an initial simplified textured mesh, we apply TQEM
to the final textured mesh obtained after both
nvdiffrec stages.
We then refine the TQEM output for 10,000 iterations
at $2048^2$ resolution, following the original ADS
iteration count and resolution, using the final
textured mesh as the reference.

\FloatBarrier                     %
\section{Implementation Details}
\label{app:impl_details}

We provide additional implementation details of
the forward simplification in
Sec.~\ref{sec:method:forward} and the closed-form
derivatives underlying the backward formulation
in Sec.~\ref{sec:method:parallel}.

\subsection{Cost Threshold and Budget Control}

We initialize the pass-wise cost threshold $\tau_p$
(Sec.~\ref{sec:method:forward}) at $\tau_1=10^{-4}$
and set $\tau_{p+1}=\sqrt{10}\,\tau_p$.
Within each pass, parallel collapse rounds continue
until no edge chosen by independent-region selection
(Sec.~\ref{sec:prelim_parallel})
satisfies $C_e\le\tau_p$.
Let $|V|$ be the current vertex count.
If a round selects more than $|V|-K$ edges,
we collapse only the $|V|-K$ lowest-cost candidates
so that the target budget $K$ is not overshot.

\subsection{Placement Branches and Fallbacks}
\label{app:impl:placement}

Consider a collapse $e=(i,j)\in\mathcal E_r$ within a pass.
We omit round superscripts when unambiguous:
$v_i,v_j$ denote the positions
$v_i^{(r-1)},v_j^{(r-1)}$ before the collapse,
and $Q_e,v_e$ denote $Q_e^{(r)},v_e^{(r)}$.
As in Eq.~\eqref{eq:qem_merge},
$Q_e=Q_i^{(r-1)}+Q_j^{(r-1)}$
has blocks $A_e,b_e,c_e$.

\paragraph{Regularized QEM placement.}
We select the regularized QEM branch when
\begin{equation}
    \frac{\mu_{\min}(A_e)}{\mu_{\max}(A_e)}>\kappa,
    \qquad \kappa=10^{-3},
    \label{eq:app:qem_admission}
\end{equation}
where $\mu_{\min}$ and $\mu_{\max}$ are the extremal
eigenvalues of $A_e$; we additionally require
$\mu_{\max}(A_e)>0$.
The merged vertex is then placed at
\begin{equation}
    v_e=-(A_e+\lambda_e I)^{-1}b_e,\qquad
    \lambda_e=\eta\max_k|(A_e)_{kk}|,\quad \eta=10^{-6}.
    \label{eq:app:qem_placement}
\end{equation}

\paragraph{Edge-segment fallback.}
When the regularized QEM branch is not selected,
we instead minimize the collapse error $E_e$
(Sec.~\ref{sec:prelim_qem}) along the edge.
With $\delta=v_j-v_i$, segment placement is used when
\begin{equation}
    \delta^\top A_e\delta>\epsilon_s,\qquad
    \epsilon_s=10^{-7},
    \label{eq:app:segment_admission}
\end{equation}
and is given by
\begin{equation}
\begin{aligned}
    \alpha^\star
        &=-\frac{\delta^\top(A_ev_i+b_e)}
                 {\delta^\top A_e\delta},\\
    \gamma_e&=\operatorname{clamp}(\alpha^\star,0,1),\\
    v_e&=v_i+\gamma_e\delta.
\end{aligned}
\label{eq:app:segment_placement}
\end{equation}
Clamping keeps $v_e$ on the edge,
at an endpoint when $\alpha^\star\notin[0,1]$.

\paragraph{Midpoint fallback.}
If Eq.~\eqref{eq:app:segment_admission} is also not satisfied,
we use $v_e=\tfrac12(v_i+v_j)$.

For each collapse, the forward pass also records the
discrete choices required by the backward pass:
the placement branch, the segment clamp state
(interior or either endpoint), and the diagonal
index $k_e$ attaining the maximum in
Eq.~\eqref{eq:app:qem_placement}.
These choices are part of the collapse history $\sigma$
(Sec.~\ref{sec:method:fixed_history})
and remain fixed during differentiation.
Across ten reconstruction samples, regularized QEM,
edge-segment, and midpoint placements account for
63.1\%, 35.6\%, and 1.3\% of in-loop collapses,
respectively.

\subsection{Gradients through Vertex Placement}
\label{app:diff:placement}
\label{app:placement_derivatives}

We derive the collapse-local VJPs for the placement
branches in Sec.~\ref{app:impl:placement},
writing $g_x$ for the gradient of the loss with respect to $x$.
Given the incoming position gradient
$g_{v_e}=g_{v_e}^{(r)}$, each branch produces
direct position gradients $g_{v_i},g_{v_j}$
and a quadric gradient
$G_e=(\partial v_e/\partial Q_e)^\top g_{v_e}$.
Writing $G_A$ and $G_b$ for the gradients with respect
to the independent blocks $A_e$ and $b_e$,
we represent the symmetric quadric gradient as
\begin{equation}
    G_e=
    \begin{bmatrix}
        G_A & \tfrac12 G_b\\
        \tfrac12 G_b^\top & 0
    \end{bmatrix}.
    \label{eq:app:local_quadric_grad}
\end{equation}
The factors $\tfrac12$ account for the two symmetric
occurrences of $b_e$ in $Q_e$, and the zero block reflects
that $c_e$ does not affect the placement.
Let $\operatorname{sym}(M)=\tfrac12(M+M^\top)$
denote the symmetric part of a matrix.

\paragraph{Regularized QEM placement.}
On this branch, Eq.~\eqref{eq:app:qem_admission} and
$\mu_{\max}(A_e)>0$ imply $A_e\succ0$.
Its diagonal entries are therefore positive,
and the absolute value in $\lambda_e$ can be dropped.
Differentiating $(A_e+\lambda_e I)v_e+b_e=0$
with the recorded $k_e$ fixed gives
\begin{equation}
    dv_e=-(A_e+\lambda_e I)^{-1}
        (dA_e\,v_e+d\lambda_e\,v_e+db_e),
    \qquad
    d\lambda_e=\eta\,(dA_e)_{k_ek_e}.
\end{equation}
With $u=(A_e+\lambda_e I)^{-1}g_{v_e}$, the VJP is
\begin{equation}
\begin{aligned}
    G_A&=-\operatorname{sym}(uv_e^\top)
         -\eta\,(u^\top v_e)\,
         \mathbf e_{k_e}\mathbf e_{k_e}^\top,\\
    G_b&=-u,\qquad g_{v_i}=g_{v_j}=0,
\end{aligned}
\label{eq:app:qem_vjp}
\end{equation}
where $\mathbf e_{k_e}$ is the $k_e$-th unit vector.
This branch propagates the placement gradient entirely
through the quadrics, including the dependence of
$\lambda_e$ on $A_e$.

\paragraph{Edge-segment placement.}
For an interior solution $\gamma_e=\alpha^\star\in(0,1)$,
let $\rho=\delta^\top A_e\delta$, $z_e=A_ev_e+b_e$, and
\begin{equation}
    \beta_e=-\frac{\delta^\top g_{v_e}}{\rho},\qquad
    h_e=g_{v_e}+\beta_e A_e\delta.
\end{equation}
Differentiating Eq.~\eqref{eq:app:segment_placement} gives
\begin{equation}
\begin{aligned}
    g_{v_i}&=(1-\alpha^\star)\,h_e-\beta_e z_e,
        &g_{v_j}&=\alpha^\star h_e+\beta_e z_e,\\
    G_A&=\beta_e\operatorname{sym}(\delta v_e^\top),
        &G_b&=\beta_e\delta.
\end{aligned}
\label{eq:app:segment_vjp}
\end{equation}
If the solution is clamped to an endpoint,
$v_e$ coincides with that endpoint and does not depend on $Q_e$:
\begin{equation}
    (g_{v_i},g_{v_j})=
    \begin{cases}
        (g_{v_e},0), & \gamma_e=0,\\
        (0,g_{v_e}), & \gamma_e=1,
    \end{cases}
    \qquad G_e=0.
    \label{eq:app:clamp_vjp}
\end{equation}

\paragraph{Midpoint placement.}
For $v_e=\tfrac12(v_i+v_j)$,
\begin{equation}
    g_{v_i}=g_{v_j}=\tfrac12 g_{v_e},\qquad G_e=0.
    \label{eq:app:midpoint_vjp}
\end{equation}

In all branches, $G_e$ is the placement term of
Eq.~\eqref{eq:quadric_merge_gradient} and is added to the
quadric gradients of both endpoints.
The direct gradients $g_{v_i}$ and $g_{v_j}$ form the
position part of the round-wise VJP $g^{(r-1)}=J_r^\top g^{(r)}$.

\subsection{Face-Quadric Derivatives}
\label{app:diff:quadric}

At the beginning of a pass, let
$f=(i,j,k)\in F_{p-1}$ be a nondegenerate face.
We write $v_i,v_j,v_k$ for its pass-input positions
$v_i^{(0)},v_j^{(0)},v_k^{(0)}$ and define
\begin{equation}
\begin{gathered}
    s=(v_j-v_i)\times(v_k-v_i),\qquad n_f=s/\|s\|,\\
    d_f=-n_f^\top v_i,\qquad
    p_f=\begin{bmatrix}n_f\\ d_f\end{bmatrix},\qquad
    w_f=\frac{\|s\|}{2\bar a},\qquad
    K_f=w_fp_fp_f^\top,
\end{gathered}
\end{equation}
so that $w_f=a_f/\bar a$ as in Sec.~\ref{sec:method:forward}.
From Eq.~\eqref{eq:initial_vertex_gradient},
the incoming face-quadric gradient is
$g_{K_f}=g_{Q_i}^{(0)}+g_{Q_j}^{(0)}+g_{Q_k}^{(0)}$.
Differentiating $K_f$ gives
\begin{equation}
    g_{p_f}=2w_f\operatorname{sym}(g_{K_f})p_f,
    \qquad
    g_{w_f}=p_f^\top g_{K_f}p_f.
\end{equation}
We partition $g_{p_f}=(g_{n_f}^\top,g_{d_f})^\top$
into normal and offset parts.
Since $d_f$ also depends on $n_f$, the total gradient on $n_f$ is
$\widetilde g_{n_f}=g_{n_f}-g_{d_f}v_i$,
and the gradient with respect to $s$ is
\begin{equation}
    g_s=\frac{1}{\|s\|}(I-n_fn_f^\top)\widetilde g_{n_f}
        +\frac{g_{w_f}}{2\bar a}n_f.
\end{equation}
The face contributions to the vertex-position
gradients follow from the cross product:
\begin{equation}
\begin{aligned}
    g_{v_j}^{f}&=(v_k-v_i)\times g_s,\\
    g_{v_k}^{f}&=g_s\times(v_j-v_i),\\
    g_{v_i}^{f}&=-g_{d_f}n_f-g_{v_j}^{f}-g_{v_k}^{f}.
\end{aligned}
\label{eq:app:face_vjp}
\end{equation}
Summing these contributions over incident faces
and adding the direct position gradients $g_{v_i}^{(0)}$
yields the pass-input gradient in
Eq.~\eqref{eq:initial_vertex_gradient}.
As in Sec.~\ref{sec:method:forward}, $\bar a$ is held
fixed during differentiation. Under a fixed collapse
history, uniformly scaling all quadrics leaves the
vertex placement unchanged in every branch.
\FloatBarrier                     %

\section{Additional Method Analysis}
\label{app:experiments}

We further analyze the forward simplifier, the fixed-history
backward pass, and iterative optimization.
Unless stated otherwise, we follow the low-poly mesh
reconstruction setup of Sec.~\ref{subsec:low-poly}
and use the same 79 shapes.
We report squared Chamfer distance (CD) scaled by $10^5$.

\subsection{Forward Simplifier}
\label{app:exp:forward}

\subsubsection{Component Ablation}
We evaluate the contributions of area-weighted face quadrics,
the cost-threshold schedule, and pass-wise quadric recomputation
by removing each component from our forward simplifier.
All variants simplify the original meshes of the
same 79 shapes (33,228 vertices on average)
to $K=1000$ without optimization.
Base denotes our parallel QEM-based edge-collapse
simplifier without these three improvements.
As shown in Table~\ref{tab:app:forward_loo}, Base yields
higher mean CD (1.879 vs.\ 1.774) and lower F1
(0.196 vs.\ 0.215) than sequential QEM, reflecting
the quality--efficiency trade-off of parallel edge collapse.
Our complete forward simplifier reduces mean CD to 1.250,
while removing any component increases CD and decreases F1.
Fig.~\ref{fig:app:forward_loo} illustrates the qualitative
effects of these components.

\begin{table}[h]
\centering
\small
\setlength{\tabcolsep}{6pt}
\caption{Component ablation of our forward simplifier
on 79 original meshes at $K=1000$, without optimization.
Metrics as in Table~\ref{tab:recon_main}. CD is scaled by $10^5$.}
\label{tab:app:forward_loo}
\begin{tabular}{lccc}
\toprule
Variant
& CD$\downarrow$
& F1$\uparrow$
& NC$\uparrow$ \\
\midrule
QEM
& 1.774 & 0.215 & 0.964 \\
Base
& 1.879 & 0.196 & 0.965 \\
Ours (w/o area-weighted face quadrics)
& 1.601 & 0.196 & \textbf{0.970} \\
Ours (w/o threshold schedule)
& 1.698 & 0.216 & 0.965 \\
Ours (w/o quadric recomputation)
& 1.505 & 0.217 & 0.966 \\
\midrule
Ours
& \textbf{1.250} & \textbf{0.230} & \textbf{0.970} \\
\bottomrule
\end{tabular}
\end{table}

\begin{figure}[h]
\centering
\includegraphics[width=\linewidth]{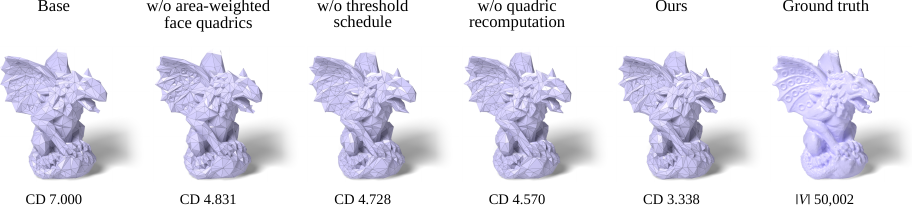}
\caption{Component ablation of the forward simplifier
on the original mesh at $K=1000$,
without optimization (Table~\ref{tab:app:forward_loo}).
CD ($\times10^{-5}$) is shown beneath each simplified
output, and $|V|$ beneath the original mesh (Ground truth).}
\label{fig:app:forward_loo}
\label{fig:app:forward_loo}
\end{figure}

\subsubsection{Comparison with Other Simplification Operators}
Table~\ref{tab:app:gt_operators} compares our forward simplifier
with QEM, vertex clustering, and PaMO on the same
79 original meshes at $K=1000$ without optimization.
These methods exhibit different trade-offs:
vertex clustering is fast but introduces substantial
geometric error due to spatial quantization, while
PaMO ensures intersection-free outputs at higher
computational cost. Sequential QEM achieves better
geometric fidelity than both but requires considerably
more time than ours.
Our forward simplifier achieves the lowest CD,
the highest F1 and NC, and the shortest simplification
time (7.5\,ms) among the evaluated methods, while
maintaining SI and SR at levels comparable to QEM.
This balance of approximation quality and efficiency
makes it suitable for repeated application within
gradient-based optimization.

\begin{table}[h]
\centering
\small
\setlength{\tabcolsep}{4pt}
\caption{Comparison of simplification operators on
79 original meshes at $K=1000$, without optimization.
QEM and vertex clustering run on CPU, whereas
PaMO and ours run on GPU. CD is scaled by $10^5$.}
\label{tab:app:gt_operators}
\begin{tabular}{lcccccc}
\toprule
Operator & CD$\downarrow$
& F1$\uparrow$ & NC$\uparrow$
& SI(\%)$\downarrow$ & SR(\%)$\downarrow$
& Time (ms)$\downarrow$ \\
\midrule
QEM
& 1.774 & 0.215 & 0.964
& 0.42 & 0.02 & 368.3 \\
Vert. Clust.
& 39.528 & 0.072 & 0.895
& 9.66 & 0.03 & 11.7 \\
PaMO
& 5.241 & 0.114 & 0.950
& \textbf{0.00} & \textbf{0.00} & 817.4 \\
\midrule
Ours
& \textbf{1.250} & \textbf{0.230} & \textbf{0.970}
& 0.47 & 0.06 & \textbf{7.5} \\
\bottomrule
\end{tabular}
\end{table}

\subsection{Backward}
\label{app:exp:gradient}

\subsubsection{Pass Count and Quadric Reduction-Forest Efficiency}

Our factorized backward substantially reduces
computation within each pass by propagating quadric
gradients through the reduction forest, allowing
face quadrics to be differentiated only once per pass
across multiple collapse rounds.
As shown in Table~\ref{tab:app:passes}, even at $K=250$,
simplification requires only 4.3 passes on average
despite 101.3 collapse rounds.
In contrast, recomputing quadrics after every round
requires repeated face-quadric differentiation.
Our pass-wise approach consequently achieves
5.5--8.8$\times$ faster backward execution,
demonstrating the efficiency of reusing quadrics
across collapse rounds.
Note that the every-round variant still computes
collapse-local VJPs in parallel, as described in
Sec.~\ref{sec:method:parallel}.

\begin{table}[h]
\centering
\small
\setlength{\tabcolsep}{6pt}
\caption{Threshold passes, collapse rounds, and operator
runtime across vertex budgets on an RTX~5090.
Measurements are averaged over 79 original meshes
without optimization, using one forward and backward
evaluation per mesh.
Passes and Rounds correspond to ours, with rounds
summed across all passes.
Every round recomputes quadrics after each collapse round.}
\label{tab:app:passes}
\begin{tabular}{rccrrrr}
\toprule
& & & \multicolumn{2}{c}{Forward (ms)}
& \multicolumn{2}{c}{Backward (ms)} \\
\cmidrule(lr){4-5}\cmidrule(lr){6-7}
$K$ & Passes & Rounds
& Ours & Every round
& Ours & Every round \\
\midrule
2000 & 1.9 & 57.5
& \textbf{6.40} & 31.24
& \textbf{0.84} & 4.59 \\
1000 & 2.6 & 74.2
& \textbf{7.48} & 42.34
& \textbf{0.96} & 6.16 \\
250 & 4.3 & 101.3
& \textbf{10.28} & 65.20
& \textbf{1.15} & 10.17 \\
\bottomrule
\end{tabular}
\end{table}

\subsubsection{Backward Work}

The quadric reduction-forest formulation in
Sec.~\ref{sec:method:parallel} performs constant-size
updates per collapse and differentiates each
pass-input face once, yielding work proportional to
$W=\sum_p\bigl(\sum_{r=1}^{R_p}|\mathcal E_r|
+|F_{p-1}|\bigr)$.
Here, $R_p$ denotes the number of collapse rounds
in pass $p$, and the pass index of $\mathcal E_r$
is omitted for brevity.
As shown in Table~\ref{tab:app:bwd_work}, our launched
GPU threads remain near 2.05--2.25 per unit of $W$
across the tested input sizes.
In contrast, an autograd replay of the same recorded
history launches 501--772 threads per unit of $W$,
while the normalized direct face-derivative count
$\sum_e|\mathcal F_e|/W$ increases from 5.40 to 21.95.
This increase reflects repeated differentiation of
the same face-quadric contributions accumulated in
vertex quadrics across successive collapses.
These results support the linear-work structure
of our factorized backward and demonstrate how
the quadric reduction forest avoids redundant
face-level differentiation across collapse rounds.

\begin{table}[h]
\centering
\small
\caption{Backward work on Gaussian random-field meshes
at $K=1000$.
Launched GPU threads and face-derivative evaluations
are normalized by $W$.
$\mathcal F_e$ denotes the faces contributing to
the vertex quadrics used to collapse edge $e$,
with the sum taken over all recorded collapses.}
\label{tab:app:bwd_work}
\begin{tabular}{rrrrccc}
\toprule
& & & & \multicolumn{2}{c}{Threads$/W$} & \\
\cmidrule(lr){5-6}
$|V_{\mathrm{in}}|$ & Passes & $\sum_p R_p$ & $W$
& Ours & Autograd & $\sum_e|\mathcal F_e|/W$ \\
\midrule
50,145    & 4 & 86  & 178,573    & 2.25 & 501 & 5.40 \\
502,141   & 4 & 146 & 1,553,225  & 2.08 & 658 & 12.25 \\
5,045,762 & 4 & 225 & 15,212,320 & 2.05 & 772 & 21.95 \\
\bottomrule
\end{tabular}
\end{table}

\subsubsection{Gradient Correctness}

We validate the factorized CUDA backward pass of
\methodname against a PyTorch autograd replay of
the same recorded collapse history.
We randomly select ten of the 79 original meshes
and perform a single forward and backward evaluation
per mesh at $K=1000$, without optimization.
For each mesh, we define a random linear loss on
the simplified vertex positions.
Let $G$ and $G_{\mathrm{AD64}}$ denote the input-vertex
gradients computed by our double-precision CUDA backward
and double-precision PyTorch autograd, respectively.
We measure the relative $L_2$ error
$\|G-G_{\mathrm{AD64}}\|_2/\|G_{\mathrm{AD64}}\|_2$
over flattened vertex gradients.
The mean and maximum relative $L_2$ errors are
$4.7\times10^{-12}$ and $2.5\times10^{-11}$,
respectively, demonstrating near-exact numerical
agreement between our CUDA backward and the autograd reference.

\subsection{Iterative Optimization}
\label{app:exp:alternatives}

\subsubsection{Training Loss Curves}
\begin{figure}[h]
\centering
\includegraphics[width=0.88\linewidth]{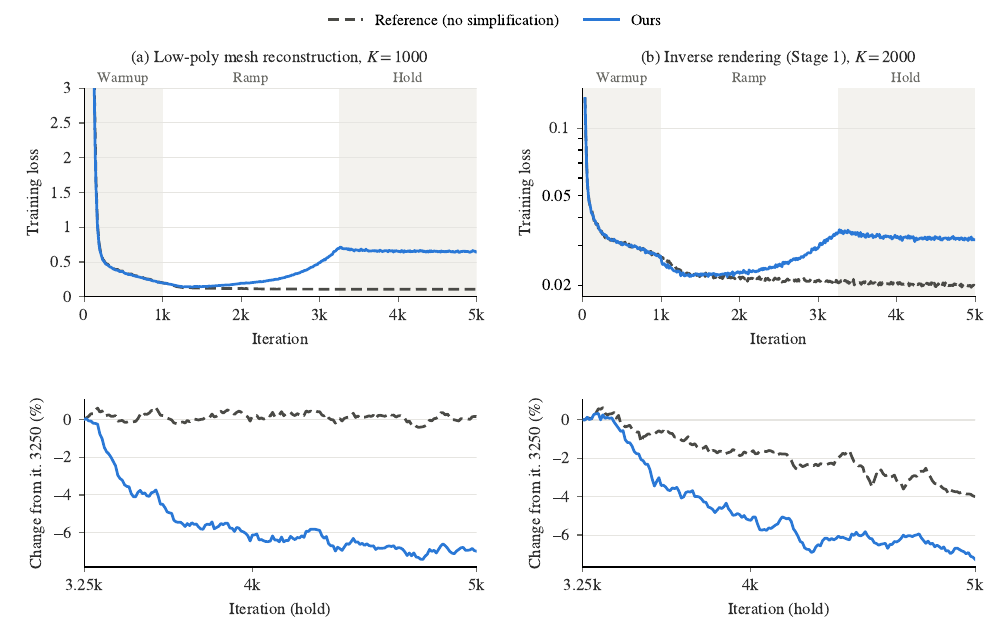}
\caption{Training loss curves of the unsimplified
reference and \methodname, aggregated using the
geometric mean across shapes or scenes.
(a) Low-poly mesh reconstruction on 79 shapes at
$K=1000$ (Table~\ref{tab:recon_main}).
(b) Stage~1 of nvdiffrec on eight NeRF-synthetic
scenes at $K=2000$ (Table~\ref{tab:inverse_rendering}).
Top: Our loss is evaluated at the current vertex budget.
Bottom: Relative loss change from iteration 3250
during the hold phase.}
\label{fig:app:losscurve}
\end{figure}

Fig.~\ref{fig:app:losscurve} compares the training losses
of \methodname and the unsimplified reference in
low-poly mesh reconstruction (Sec.~\ref{subsec:low-poly})
and inverse rendering (Sec.~\ref{sec:exp:inverse_rendering}).
With $T=5000$, the warmup ($t_w=0.2T$) and
budget-reduction ($t_r=0.45T$) schedules divide
training into warmup (0--1000), ramp (1000--3250),
and hold (3250--5000) phases.
Both methods optimize the same dense geometry during
warmup. During the ramp, our training loss increases
as the vertex budget progressively decreases.
Once the target budget is reached, our training loss
continues to decrease throughout the hold phase,
exhibiting a clear convergence trend despite
repeated simplification during optimization.

\subsubsection{Post-Simplification Fine-Tuning}

An alternative is to simplify the dense reconstruction
once and then optimize the simplified mesh with fixed
connectivity.
To construct this baseline, we apply our forward
simplifier as post-processing and refine the resulting
vertex positions for 2,000 additional iterations using Large Steps with its multi-view
reconstruction objective described in
Sec.~\ref{app:exp:low-poly}.
Table~\ref{tab:app:pft} shows that, even after these
additional 2,000 iterations, post-simplification fine-tuning
fails to match the in-loop output of \methodname
in both CD and F1 across all tested vertex budgets.
The gaps in both metrics increase under tighter budgets,
with the largest differences at $K=500$.

\begin{table}[h]
\centering
\small
\setlength{\tabcolsep}{4pt}
\caption{Post-simplification fine-tuning on 79 shapes.
Init denotes the base FlexiCubes reconstruction
simplified by our forward simplifier as post-processing.
Post-FT denotes 2,000 additional fine-tuning iterations
using Large Steps.
CD ($\times10^{-5}$) and F1 evaluate the final simplified meshes.}
\label{tab:app:pft}
\begin{tabular}{rcccccc}
\toprule
& \multicolumn{3}{c}{CD$\downarrow$}
& \multicolumn{3}{c}{F1$\uparrow$} \\
\cmidrule(lr){2-4}\cmidrule(lr){5-7}
$K$ & Init & Post-FT & Ours
& Init & Post-FT & Ours \\
\midrule
2000 & 0.748 & 0.718 & \textbf{0.704}
     & 0.314 & 0.333 & \textbf{0.339} \\
1000 & 1.528 & 1.351 & \textbf{1.259}
     & 0.230 & 0.254 & \textbf{0.261} \\
500  & 3.971 & 3.664 & \textbf{2.891}
     & 0.156 & 0.172 & \textbf{0.182} \\
\bottomrule
\end{tabular}
\end{table}

\subsubsection{Progressive Budget Schedule Ablation}

\begin{table}[!t]
\centering
\small
\setlength{\tabcolsep}{5pt}
\caption{Budget schedule ablation on 79 FlexiCubes
shapes at $K=1000$.
CD ($\times10^{-5}$), F1, and SI are evaluated on the
largest connected component (LCC).
Failures count shapes with either a broken surface
or excessive interior shells relative to
QEM applied to the same dense reconstruction.
Pre-existing interior shells inherited from the
unsimplified FlexiCubes reconstruction are excluded
from failure counts.
The common-success subset contains the 50 shapes
where all three schedules succeed.}
\label{tab:app:schedule}
\begin{tabular}{lccccc}
\toprule
Schedule & CD$_{\mathrm{mean}}\downarrow$
& CD$_{\mathrm{med}}\downarrow$
& F1$\uparrow$ & SI(\%)$\downarrow$
& Failures$\downarrow$ \\
\midrule
\multicolumn{6}{l}{\textit{All 79 shapes}} \\
\midrule
w/o warmup ($t_w=0$)
& 1,321.681 & 0.745 & 0.262 & 1.52 & 25 \\
w/o ramp ($t_r=0$)
& 528.277 & \textbf{0.742} & \textbf{0.266}
& 4.37 & 8 \\
Ours
& \textbf{1.259} & 0.757 & 0.261
& \textbf{0.62} & \textbf{0} \\
\midrule
\multicolumn{6}{l}{\textit{Common-success subset}} \\
\midrule
w/o warmup ($t_w=0$)
& 1.254 & \textbf{0.524} & 0.347 & 0.99 & -- \\
w/o ramp ($t_r=0$)
& 1.115 & 0.526 & \textbf{0.350} & 1.03 & -- \\
Ours
& \textbf{1.065} & 0.532 & 0.343
& \textbf{0.77} & -- \\
\bottomrule
\end{tabular}
\end{table}

\begin{figure}[!t]
\centering
\includegraphics[width=0.5\linewidth]{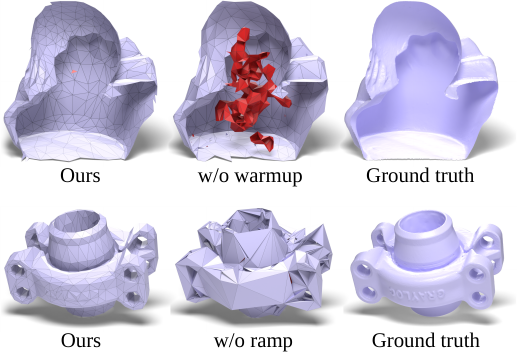}
\caption{Representative failure cases in the budget
schedule ablation at $K=1000$
(Table~\ref{tab:app:schedule}).
Top: An interior shell remains connected to the
exterior surface and is therefore retained in the
largest connected component, as shown in the cutaway.
Bottom: The reconstructed surface breaks under
abrupt budget reduction.
Red highlights interior faces.}
\label{fig:app:schedule}
\end{figure}

We ablate the progressive vertex-budget schedule
introduced in Sec.~\ref{sec:method:scheduling}
by removing either the initial optimization without
simplification (warmup, $t_w=0$) or the progressive
vertex-budget reduction (ramp, $t_r=0$).
We follow the FlexiCubes reconstruction setting
of Table~\ref{tab:recon_main} at $K=1000$ and define
failures relative to QEM post-processing applied
to the same base reconstruction, based on severe
surface degradation or excessive interior shell artifacts.
Table~\ref{tab:app:schedule} reports 25 failures without
warmup and 8 without the ramp, compared with none
under our complete schedule.
Early in optimization, randomly initialized SDFs
can produce highly fragmented meshes with interior
shell artifacts.
Such geometry is problematic for simplification:
edge collapse cannot merge disconnected components,
and interior shells connected to the exterior surface
cannot be removed by largest-component filtering.
These observations motivate the warmup phase,
allowing the upstream geometry to become more
coherent before simplification begins.
Without the ramp, imposing the target budget abruptly
after warmup leads to broken surfaces and increases
SI from 0.62\% to 4.37\% (7.0$\times$).
While median CD and F1 remain comparable across
schedules, our complete schedule avoids the severe
geometric failures observed in both ablations.
Fig.~\ref{fig:app:schedule} illustrates representative
failure cases.

\FloatBarrier                     %
\section{Additional Inverse Rendering Results}
\label{app:exp:tasks}

\subsection{Detailed nvdiffrec Results}
\begin{figure}[p]
\centering
\includegraphics[width=\linewidth,height=0.86\textheight,keepaspectratio]{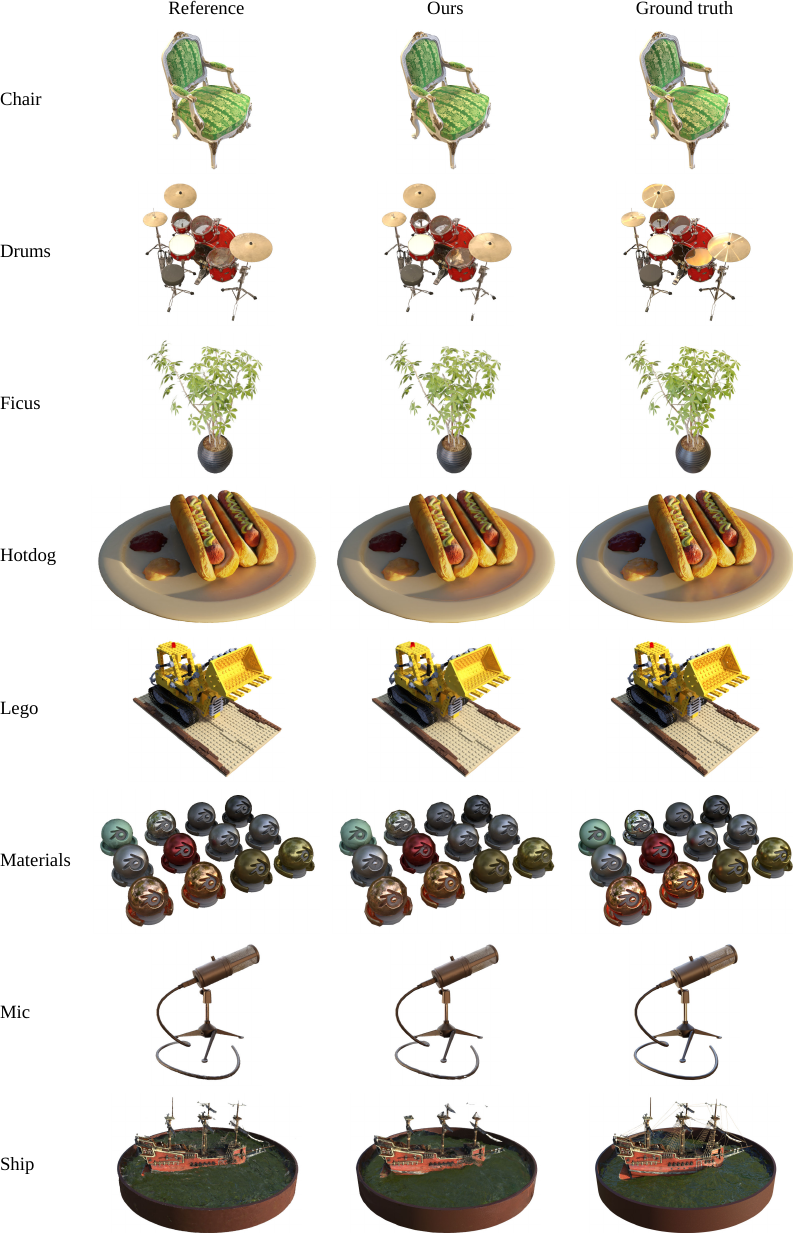}
\caption{Qualitative comparison of inverse rendering
with nvdiffrec on the eight NeRF-synthetic scenes
after the full run at $K=2000$
(Table~\ref{tab:inverse_rendering}).
Test views rendered from the reference
and ours, alongside ground-truth images.
The average vertex counts are 51{,}512 for the
reference and 2{,}000 for ours.
Reference denotes the unsimplified reconstruction from
the base nvdiffrec pipeline.
}
\label{fig:app:irender_all}
\end{figure}

Table~\ref{tab:app:nvdr_scene} reports per-scene PSNR for the
inverse-rendering experiments with nvdiffrec in
Table~\ref{tab:inverse_rendering} at $K=2000$.
Reference denotes the unsimplified reconstruction from
the base nvdiffrec pipeline.
Fig.~\ref{fig:app:irender_all} shows full-image renderings
of Reference and ours alongside ground-truth images
across the eight NeRF-synthetic scenes.

\begin{table}[h]
\centering
\small

\caption{Novel-view PSNR (dB) per NeRF-synthetic scene
at $K=2000$, corresponding to
Table~\ref{tab:inverse_rendering}.
Reference denotes the unsimplified reconstruction
from the base nvdiffrec pipeline.
Methods marked $^*$ simplify the final textured
reference; their Stage 1 values in Table~\ref{tab:inverse_rendering}
are inherited from Reference and omitted here.}

\label{tab:app:nvdr_scene}
\begin{tabular}{lccccccccc}
\toprule
Method & Chair & Drums & Ficus & Hotdog & Lego & Mat. & Mic & Ship & Mean \\
\midrule
\multicolumn{10}{l}{\emph{Stage 1}} \\
Reference & 28.97 & 23.18 & 27.93 & 30.44 & 26.52 & 26.04 & 28.58 & 23.35 & 26.88 \\
Low-res & 27.35 & 19.68 & 21.68 & 27.41 & 22.06 & 22.78 & 27.18 & 21.72 & 23.73 \\
QEM & 28.18 & 19.65 & 20.14 & 29.98 & 21.16 & 21.89 & 28.12 & 20.69 & 23.73 \\
Ours (fwd. only) & 28.60 & 20.47 & 19.63 & 30.10 & 22.97 & 22.18 & 28.38 & 21.11 & 24.18 \\
\midrule
Ours & \textbf{28.73} & \textbf{21.41} & \textbf{23.29} & \textbf{30.56} & \textbf{24.65} & \textbf{24.21} & \textbf{28.51} & \textbf{23.13} & \textbf{25.56} \\
\midrule
\multicolumn{10}{l}{\emph{Full run}} \\
Reference & 31.32 & 24.59 & 30.96 & 32.70 & 29.24 & 27.19 & 30.47 & 25.71 & 29.02 \\
TQEM$^*$ & 24.90 & 19.06 & 21.08 & 27.71 & 19.46 & 20.59 & 27.45 & 19.53 & 22.47 \\
ADS$^*$ & 29.09 & 21.99 & 24.54 & 30.49 & 23.34 & 23.62 & 29.42 & 23.18 & 25.71 \\
Low-res & 29.60 & 20.67 & 24.26 & 29.37 & 22.80 & 24.71 & 29.38 & 23.66 & 25.56 \\
QEM & 30.34 & 20.52 & 22.39 & 31.31 & 22.49 & 24.16 & 30.23 & 21.97 & 25.43 \\
Ours (fwd. only) & \textbf{30.72} & 22.12 & 23.29 & 31.35& 23.28 & 24.50 & 30.20 & 22.81 & 26.03 \\
\midrule
Ours & 30.42 & \textbf{22.79} & \textbf{27.68} & \textbf{31.55} & \textbf{25.68} & \textbf{25.24} & \textbf{30.40} & \textbf{24.56} & \textbf{27.29} \\
\bottomrule
\end{tabular}
\end{table}

\subsection{Scene Reconstruction with ExMesh}

To evaluate the generality of \methodname across
inverse-rendering pipelines, we integrate it into
ExMesh and evaluate 15 scans from the DTU dataset
at $K=1000$, using the authors' released implementation
and experimental settings.
ExMesh jointly optimizes explicit mesh geometry
and UV textures with adaptive topology updates.
We retain its initialization, loss functions,
per-scan loss weights, and 10,000-iteration optimization.
We follow the same progressive vertex-budget schedule
as in Sec.~\ref{sec:experiments}, with
$t_w=0.2T$ and $t_r=0.45T$.
Following ExMesh's atlas-rebuilding schedule,
we rebuild the simplified mesh's UV atlas at
iterations 2000, 4000, and 6000, and once
after training. Between these updates, UV
correspondence is maintained through
closest-surface texture mapping.
Unlike nvdiffrec, ExMesh has no separate
texture-refinement stage after simplification.
We use the final dense textured mesh from the
unmodified ExMesh pipeline as Reference.
We therefore compare against TQEM and ADS as
texture-aware post-processing baselines, both
applied to this Reference.
For ADS, we follow the refinement settings in
Sec.~\ref{appendix:inverse_rendering_setup}.

\begin{figure}[p]
\centering
\includegraphics[width=\linewidth,height=0.86\textheight,keepaspectratio]{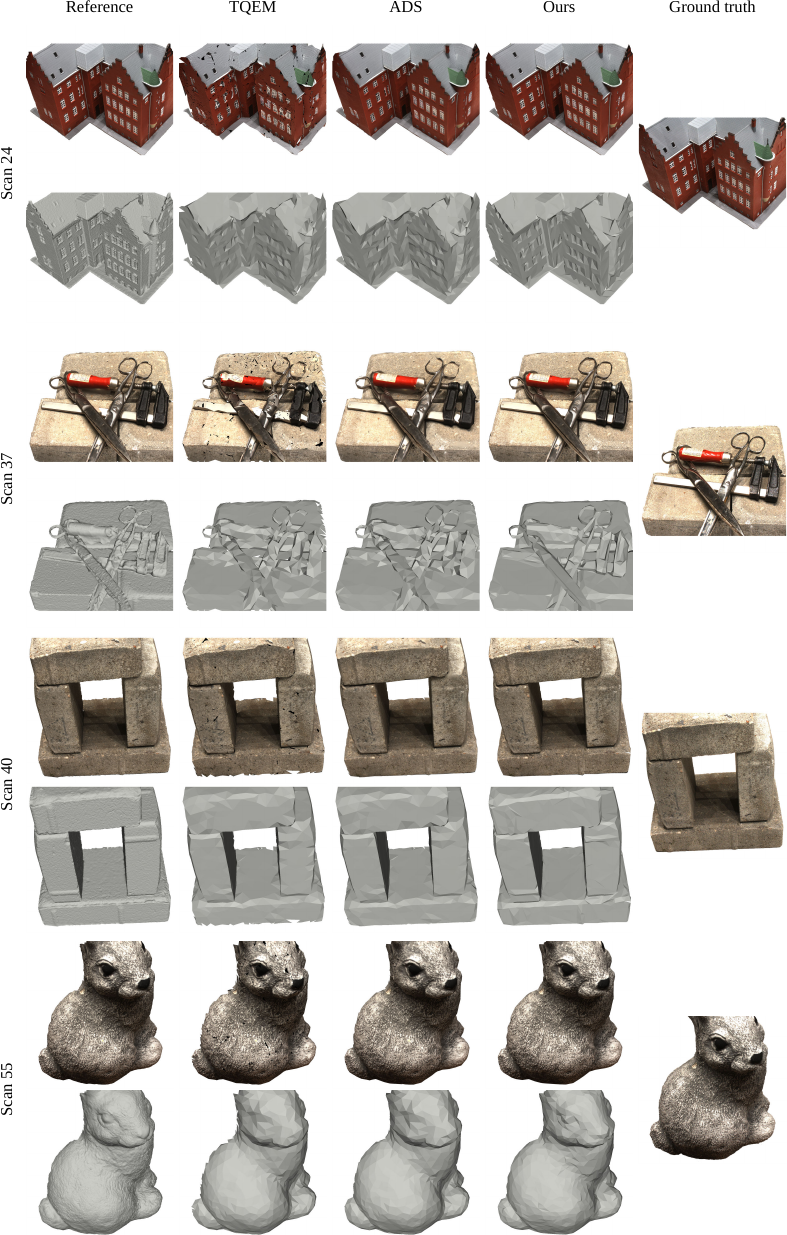}

\caption{Qualitative comparison of scene reconstruction
with ExMesh on scans 24, 37, 40, and 55 from the
DTU dataset at $K=1000$ (Table~\ref{tab:app:exmesh}).
Reference denotes the dense textured mesh reconstructed
by the original ExMesh pipeline.
Upper rows show textured renderings, lower rows show
mesh geometry, and Ground truth shows the corresponding
input photographs.}

\label{fig:app:exmesh_dtu}

\end{figure}

As shown in Table~\ref{tab:app:exmesh}, ours achieves
the lowest Chamfer distance on all 15 scans
and the highest PSNR on 14 of 15 scans among
the simplified methods.
Fig.~\ref{fig:app:exmesh_dtu} further illustrates
better preservation of planar surfaces and
sharp geometric features.

\begin{table}[h]
\centering
\small
\setlength{\tabcolsep}{2pt}

\caption{ExMesh reconstruction on 15 DTU scans at $K=1000$.
Reference denotes the unsimplified output of the
unmodified ExMesh pipeline before post-processing
simplification. We report official DTU Chamfer distance
(mm) and PSNR (dB) on training views.
Reference meshes contain 103{,}423 vertices on average,
while all simplified meshes contain 1{,}000 vertices.}

\label{tab:app:exmesh}
\begin{tabular*}{\linewidth}{@{\extracolsep{\fill}}l*{16}{c}@{}}
\toprule
Scan & 24 & 37 & 40 & 55 & 63 & 65 & 69 & 83
& 97 & 105 & 106 & 110 & 114 & 118 & 122 & Mean \\
\midrule
\multicolumn{17}{l}{\textit{CD$\downarrow$}} \\
Reference
& 0.502 & 0.697 & 0.331 & 0.359 & 0.784
& 0.654 & 0.562 & 1.116 & 0.976 & 0.611
& 0.479 & 0.758 & 0.320 & 0.396 & 0.405 & 0.597 \\
TQEM
& 0.702 & 0.833 & 0.420 & 0.473 & 0.823
& 0.839 & 0.829 & 1.155 & 1.040 & 0.667
& 0.738 & 0.810 & 0.458 & 0.637 & 0.561 & 0.732 \\
ADS
& 0.749 & 0.815 & 0.415 & 0.431 & 0.851
& 0.809 & 0.800 & 1.217 & 1.054 & 0.704
& 0.695 & 0.824 & 0.417 & 0.596 & 0.543 & 0.728 \\
Ours
& \textbf{0.591} & \textbf{0.727} & \textbf{0.353}
& \textbf{0.413} & \textbf{0.782} & \textbf{0.734}
& \textbf{0.649} & \textbf{1.129} & \textbf{0.965}
& \textbf{0.634} & \textbf{0.602} & \textbf{0.710}
& \textbf{0.386} & \textbf{0.495} & \textbf{0.464}
& \textbf{0.642} \\
\midrule
\multicolumn{17}{l}{\textit{PSNR$\uparrow$}} \\
Reference
& 22.97 & 20.48 & 24.00 & 26.68 & 27.65
& 25.59 & 25.47 & 32.50 & 23.99 & 27.55
& 29.86 & 24.02 & 26.32 & 32.75 & 33.11 & 26.86 \\
TQEM
& 17.80 & 17.20 & 20.20 & 21.53 & 24.87
& 20.45 & 23.31 & 30.10 & 21.59 & 24.47
& 24.90 & 23.10 & 22.75 & 27.34 & 27.84 & 23.16 \\
ADS
& 21.10 & 19.43 & 22.87 & 24.73 & 27.11
& 24.12 & 24.78 & 31.95 & 23.41 & 26.86
& 27.86 & \textbf{23.76} & 25.42 & 30.60
& 31.10 & 25.67 \\
Ours
& \textbf{21.88} & \textbf{20.00} & \textbf{23.14}
& \textbf{25.23} & \textbf{27.27} & \textbf{24.86}
& \textbf{25.03} & \textbf{32.04} & \textbf{23.56}
& \textbf{26.93} & \textbf{28.42} & 23.73
& \textbf{25.77} & \textbf{31.18} & \textbf{32.00}
& \textbf{26.07} \\
\bottomrule
\end{tabular*}
\end{table}

\FloatBarrier                     %
\clearpage                       %

\section{Limitations and Future Work}
\label{app:limitations}

Although \methodname is a training-free, drop-in
differentiable mesh simplification layer, its
effectiveness depends on the quality of intermediate
meshes produced during optimization.
When upstream mesh generators produce fragmented geometry
or interior shell artifacts, applying simplification before
these defects are resolved can exacerbate these issues
and may lead to optimization failures.
Under aggressive simplification, the many-to-one
nature of the collapse map can leave parts of
the intermediate geometry weakly constrained,
potentially producing floating artifacts.

\paragraph{Future work.}
Our inverse-rendering experiments demonstrate
joint optimization of geometry and appearance.
However, UV correspondence is handled separately
from our simplifier. Extending our geometry-based
forward simplifier with UV-aware collapse costs
and seam-aware attribute propagation could reduce
reliance on separate UV processing and facilitate
integration into a broader range of textured-mesh
optimization pipelines.

\end{document}